\documentclass[preprint,pre,aps,amsfonts,amsmath,superscriptaddress,nofootinbib]{revtex4}
\usepackage{amsmath,amssymb,amscd}
\usepackage{empheq}
\usepackage{graphicx}
\usepackage{bm}
\usepackage{color}

\newcommand{\bea}{\begin{eqnarray}}
\newcommand{\eea}{\end{eqnarray}}

\def\XXint#1#2#3{{\setbox0=\hbox{$#1{#2#3}{\int}$}
     \vcenter{\hbox{$#2#3$}}\kern-.5\wd0}}

\begin{document}
%
%%%%%%%%%%%%%%%%%%%%
%
\title{Universal record statistics for random walks and L\'evy flights with a nonzero staying probability}
\author{Satya N. Majumdar}
\email{satya.majumdar@lptms.u-psud.fr}
\affiliation{LPTMS, CNRS, Univ. Paris-Sud, Universit\'e Paris-Saclay, 91405 Orsay, France.}
\author{Philippe Mounaix}
\email{philippe.mounaix@polytechnique.edu}
\affiliation{CPHT, CNRS, Ecole
Polytechnique, IP Paris, F-91128 Palaiseau, France.}
\author{Gr\'egory Schehr}
\email{gregory.schehr@lptms.u-psud.fr}
\affiliation{Sorbonne Universit\'e, Laboratoire de Physique Th\'eorique et Hautes Energies, CNRS, UMR 7589, 4 Place Jussieu, 75252 Paris Cedex 05, France}
\date{\today}
\begin{abstract}
We compute exactly the statistics of the number of records 
in a discrete-time random walk model on a line where the walker stays at a given position with a nonzero
probability $0\leq p \leq 1$, while with the complementary probability $1-p$, it jumps to a new position with a jump length drawn from a continuous and symmetric
distribution $f_0(\eta)$. We have shown that, for arbitrary $p$, the statistics of records up to step $N$ is completely universal, i.e., independent of $f_0(\eta)$ for
any $N$. We also compute the connected two-time correlation function $C_p(m_1, m_2)$ of the record-breaking events at times $m_1$ and
$m_2$ and show it is also universal for 
all $p$. Moreover, we demonstrate that $C_p(m_1, m_2)< C_0(m_1, m_2)$ for all $p>0$, indicating that a nonzero $p$ induces additional anti-correlations between record events.
We further show that these anti-correlations lead to a drastic reduction in the fluctuations of the record numbers with increasing $p$. This is manifest in the Fano factor, i.e. the ratio of the variance and the mean of the record number, which we compute explicitly. We also show that an interesting scaling limit emerges when $p \to 1$, $N \to \infty$ with the product $t = (1-p)\, N$ fixed. We compute exactly the associated universal scaling functions for the mean, variance and the Fano factor of the number of records in this scaling limit. 
\end{abstract}
%\pacs{02.50.Ga, 05.40.Fb, 05.45.Tp}
%
\maketitle
%
%%%%%%%%%%%%%%%%%%%%
%
\section{Introduction}\label{intro}

Records are ubiquitous in nature: in sports, climate science, finance, disordered systems, earthquake models, etc \cite{Cha1952,hoyt,basset,SZ1999,benestad,RP2006,WK2010,AB2010,WHK2013,records_finance,WBK2011,SL2014,records_hydrology,Gembris,sports,FWK2012,GL2008,sibani} -- for a recent review see Ref.~\cite{review_records}. Let us consider a time
series in discrete-time with $N$ entries $\{x_1, x_2, \cdots x_N\}$. This may represent the price of a stock or the daily average temperature 
at a given place as a function of the days. A record (upper one) happens at step $m$ if $x_m > \{x_1, x_2, \cdots, x_{m-1} \}$, i.e., the $m$-th entry
is bigger than all the previous entries. The most natural observable is the number of such records $R_N$ in a time series of size $N$. When the
entries are random variables drawn from some underlying distribution (either independent or correlated), $R_N$ is clearly a random variable and
studying its statistics is what is called ``record statistics''.  There have been a lot of studies of the record statistics for various models of the time
series: one of the interesting questions is how universal are the statistics of $R_N$ and also how it depends on the correlations between the 
entries of the time series \cite{review_records}.

The classical and the most well studied model \cite{Res1987,ABN1992,BG2001,Nevzorov} corresponds to the case when the 
underlying variables $x_i$'s are uncorrelated, each drawn independently from a continuous distribution 
$\varphi(x)$. We call this model IICD (independent, identically and continuously distributed random variables). 
In this case, remarkably, the average number of records (and even the higher moments) $\langle R_N \rangle$ is
completely universal for all $N$, i.e., independent of $\varphi(x)$. Indeed, it is given by the simple formula
\bea \label{av_RN}
\langle R_N \rangle = 1 + \frac{1}{2} + \cdots + \frac{1}{N} \;.
\eea
In particular, for large $N$, it grows rather slowly as $R_N \simeq \log N$. Similarly the variance of $R_N$ can also 
be computed and turns out to be universal for all $N$. In particular, for large $N$, the variance also grows slowly as, $V_N = \langle R_N^2 \rangle - \langle R_N \rangle^2 \simeq \log N$.
Thus the ratio of the variance and mean, known as the Fano factor \cite{Fano}, 
\bea \label{def_Fano}
F_N = \frac{V_N}{\langle R_N\rangle}
\eea 
approaches to unity, i.e., $F_N \to 1$ as $N \to \infty$. Let us recall that, had $R_N$ a Poissonian statistics, the Fano factor would be exactly $F_N=1$,
for all $N$. Thus the deviation from unity of the Fano factor for finite $N$ can be taken as a measure of the deviation from a Poissonian statistics.

One simple way to analyse the average number of records is by introducing the binary variable $\sigma_m$,
which takes value $\sigma_m = 1$ if a record happens at step $m$ and $\sigma_m=0$ otherwise. Quite generally, the 
number of records $R_N$ can be written as
\bea \label{rel_R_sigma}
R_N  = \sum_{m\le N} \sigma_m \;.
\eea 
Note that this is true for any time-series, independent or correlated. Taking average in \eqref{rel_R_sigma} gives
\bea \label{rel_R_sigma2}
\langle R_N \rangle =  \sum_{m\le N} \langle \sigma_m\rangle \;,
\eea 
where $\langle \sigma_m\rangle$ is the probability that a record happens at step $m$ and is usually known as the record rate. 
In the case of the IICD model, it is clear that $\langle \sigma_m \rangle = 1/m$ because the probability that the $m$-th
event is the maximum among $m$ IICD random variables is simply $1/m$ since any of
the $m$-th variables can be the maximum with equal probability. Thus $\langle R_N \rangle$, for any $N$, is universal, i.e., independent
of $\varphi(x)$. The reason why even the higher moments of $R_N$, for the IICD model,
is universal can be traced back to the fact that the record-breaking events $\sigma_m$'s turn out to be completely uncorrelated in this case, 
~i.e., \bea \label{correl_sigma_iid}
\langle \sigma_{m_1} \sigma_{m_2} \rangle = 
\begin{cases}
&\langle \sigma_{m_1} \rangle \langle \sigma_{m_2} \rangle \quad, \quad m_1 \neq m_2 \\
&\langle \sigma_{m_1} \rangle \quad, \quad \quad \quad \; \;m_1 = m_2 \;.
\end{cases}
\eea
This crucial property holds only when the entries are independent and their distribution $\varphi(x)$ is a continuous function. 
A natural question is: what happens to the correlations between the $\sigma_m$'s and the statistics of $R_N$
when the entries of the time series are still independent, but their distribution 
$\varphi(x)$ is not continuous. Recently, this question came up in the context
of the study of records in rainfall precipitation time-series where the entry $x_m$ represents the amount of rainfall
on the $m$-th day during the rainy season in a particular place \cite{MBK2019}.  In some days, there is no rainfall at
all, making it a dry day. This corresponds to having a delta-peak at $x=0$ with some probability weight $0\leq p \leq 1$
in the rainfall distribution $\varphi(x)$, i.e., 
\bea \label{delta_iid}
\varphi(x) = p \, \delta(x) + (1-p)\,\varphi_0(x) \;,
\eea
where $\varphi_0(x)$ is a continuous distribution normalized to one. How does the presence of a nonzero $p>0$ representing the delta-weight affect
the statistics of $\sigma_m$'s and that of $R_N$? This problem was recently studied in \cite{MBK2019} where it was shown
that any nonzero $p$ introduces {\it anti-correlations} among the $\sigma_m$ variables. This naturally affects the mean and the 
variance of $R_N$ and hence the Fano factor $F_N$ in \eqref{def_Fano}. As $p$ increases from $0$, the Fano factor decreases from
unity, which was shown to be a consequence of the anti-correlations in the $\sigma_m$'s. Thus the effect of increasing $p$ was to suppress
the fluctuations of the record number. These theoretical predictions were validated
by comparison with real climate data \cite{MBK2019}.

Going beyond the uncorrelated variables, records statistics for strongly correlated entries in the time series
have attracted much attention in recent years (see for example the review \cite{review_records}). In general, the presence of correlations between the entries makes
the study of record statistics for correlated variables much harder than in the uncorrelated case. However, there
exists one exactly solvable model with strong correlations where the entries $x_k$'s correspond to the successive
positions of a discrete-time random walker on a continuous line \cite{Feller,MZ2008}. Consider a random walker starting at the origin $x_0=0$,
and evolving via the Markov jump process
\bea \label{def_RW}
x_m = x_{m-1} + \eta_m
\eea
where the jump lengths $\eta_m$'s are IICD, each drawn from a symmetric and continuous distribution $f(\eta)$. This model includes
L\'evy flights where the jump distribution has a fat tail: $f(\eta) \sim |\eta|^{-1-\mu}$ for large $|\eta|$ and $0<\mu \leq 2$. Here, 
by convention, the initial position is counted as a record. What can we say about the statistics of the number of records $R_N$?
Quite remarkably the statistics of $R_N$ turns out to be again universal for all $N$, i.e., independent of the jump distribution
$f(\eta)$ \cite{MZ2008}. For example, the average number of records is given by the formula \cite{MZ2008}
\bea \label{av_RN_rw}
\langle R_N \rangle = (2N+1)\,{2N \choose N}\, 2^{-2N} \simeq \sqrt{4N/\pi} \quad, \quad {\rm as} \; N \to \infty \;.
\eea
Similarly, the variance is also universal for all $N$ and in particular, for large $N$, it grows as
\bea \label{var_RW}
V_N = \langle R_N^2 \rangle - \langle R_N \rangle^2 \simeq 2 \left( 1 - \frac{2}{\pi}\right) \, N \quad, \quad {\rm as} \; N \to \infty \;.
\eea
Consequently, the Fano factor in \eqref{def_Fano} behaves, for large $N$, as
\bea \label{Fano_RW}
F_N = \frac{V_N}{\langle R_N \rangle}\simeq \left( \sqrt{\pi} - \frac{2}{\sqrt{\pi}}\right)\,\sqrt{N} \quad, \quad {\rm as} \; N \to \infty \;,
\eea
reflecting the fact that the variance and the mean are not of the same order for large $N$. It turns out that
the mechanism responsible for this universality in the random walk model with IICD symmetric jumps can be traced
back to the celebrated Sparre Andersen theorem for random walks -- hence this mechanism is very different from that 
of the uncorrelated case \cite{MZ2008, review_records}. The robustness of this universality of the record statistics has 
been investigated recently in a number of variants of the basic random walk model~\cite{PLDW2009,Sanjib2011,MSW2012,WMS2012,GMS2015b,GMS2016,Cha15b,MMS2020,MDMS2020a,MDMS2020b,LM2020}.

\begin{figure}[t]
\includegraphics[width = 0.7\linewidth]{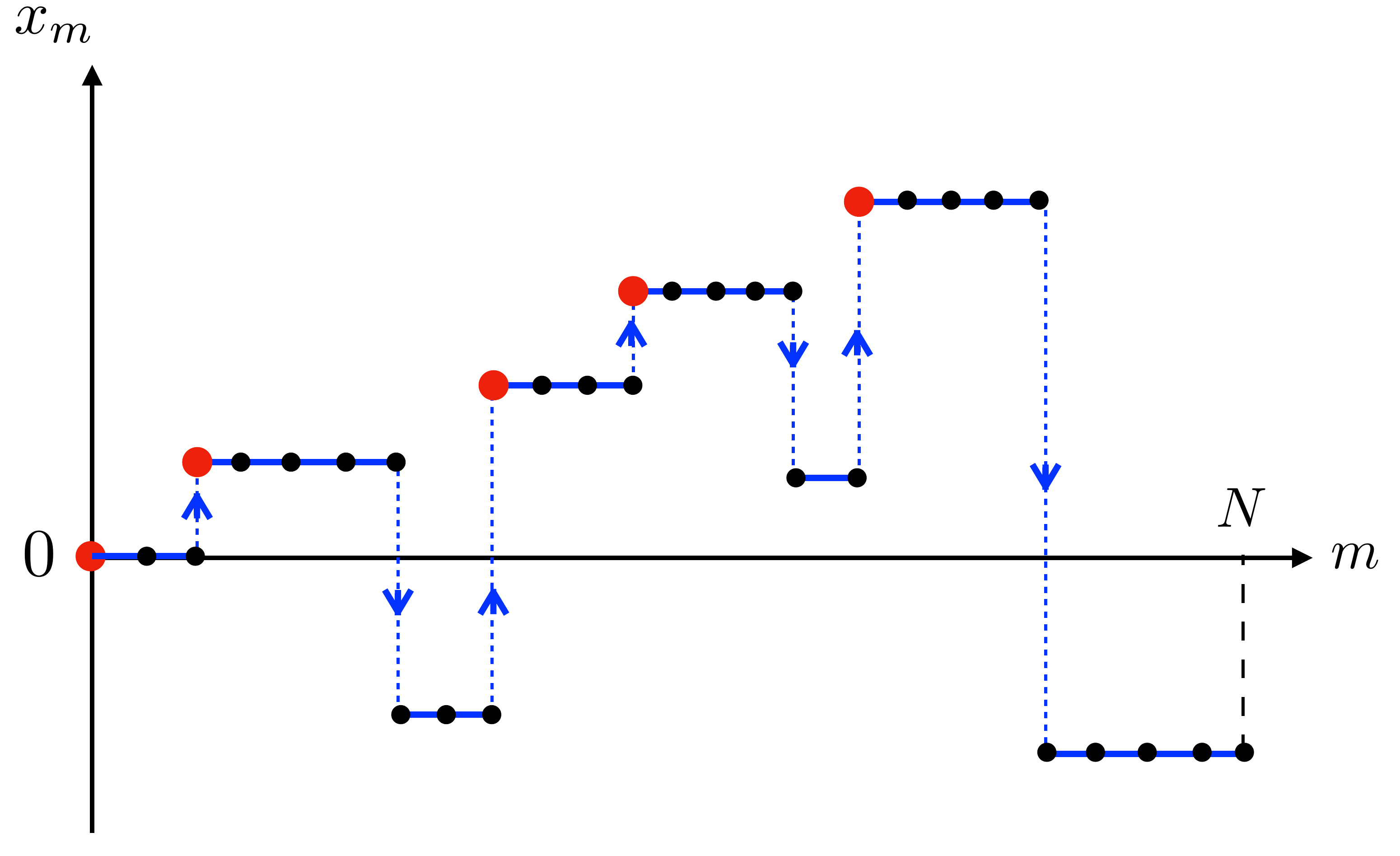}
\caption{Schematic representation of a trajectory of a discrete-time random walk of $N$ steps on the line [see Eq. (\ref{def_RW})] where the walker, stays at a given position with probability $p$ and with the complementary probability $1-p$ it jumps to a new position (as shown by arrows) with the jump length drawn from a symmetric and continuous jump distribution $f_0(\eta)$ [see Eq. (\ref{jump_distribution})]. The walker starts at the origin at step $0$, i.e., $x_0=0$. A record occurs when the random walker reaches a new maximal value for the first time, as shown by the red dots. Note that the initial position is counted as a record. Our main observable is the number of records $R_N$ up to step $N$, which is simply the number of red dots. Here $R_N =5$.}\label{Fig:traj}
\end{figure}

Following the uncorrelated model, it is then natural to ask what happens to the record statistics in this random walk
model when the jump distribution, while still symmetric, ceases to be continuous. For example, what can we say about the 
record statistics for the jump distribution with a delta peak at the origin, as in the uncorrelated model in \eqref{delta_iid}
\begin{equation}\label{jump_distribution}
f(\eta)=p\, \delta(\eta)+(1-p)\, f_0(\eta),
\end{equation}
where $0\le p\le 1$ and $f_0(\eta)$ is a continuous and symmetric distribution normalized to one?
This model naturally occurs when the walker stays, with probability $p$, at a given site and jumps 
with the remaining probability $1-p$ by a random amount $\eta$ drawn from $f_0(\eta)$. Note that, for nonzero
$p$, the entries $x_m$'s can be highly degenerate. So here we call an entry a record when its value is attained 
for the first time. If the walker attains this value at later times, those events are not counted as records. For a schematic
representation of the random walk trajectory of $N$ steps see Fig. \ref{Fig:traj} where 
the records are marked in red.

In fact this model with a jump distribution in Eq. (\ref{jump_distribution}) is just a discrete-time version
of a continuous-time model known as the ``Instantaneous Run''  (IR) model studied recently in the context
of a run-and-tumble particle \cite{MDMS2020b}. Indeed we see that in our model the waiting time distribution for the walker
at a given site is simply $(1-p)\, p^n$. Hence in the limit $p \to 1$, $n \to \infty$ but with the product $(1-p)\,n = t$ fixed,
our model reduces to a continuous time random walk (CTRW) model \cite{MS73} with an exponential waiting time distribution $e^{-t}$.
This corresponds to a walker which waits a random exponential time with mean $1$ at a given point in space and then
jumps by a random distance $\eta$ drawn from a symmetric and continuous distribution $f_0(\eta)$. Indeed this is precisely 
the IR model studied recently \cite{MDMS2020b} where the record statistics was computed exactly and was shown to be universal 
for all $t$, i.e., independent of~$f_0(\eta)$.  

In this paper, our main focus is to study the record statistics as a function of increasing $p$ in the discrete-time model, including
the continuous-time limit $p\to 1$. Our main results are twofold:

\begin{itemize}

\item[$\bullet$] {First we show that for any fixed $0\leq p \leq 1$, the statistics of the record number $R_N$ up to $N$ steps is universal 
for all $N$, i.e., independent of the distribution $f_0(\eta)$ in Eq. (\ref{jump_distribution}). We compute the mean $\langle R_N\rangle(p)$ and the variance $V_N(p)$, exactly for
all $p$ and all $N$. Our results interpolate smoothly between the two known limits $p \to 0$ (the standard RW model \cite{MZ2008}) and $p \to 1$ (the IR model \cite{MDMS2020b}).}  

\item[$\bullet$] {Our second main result is the following. We first recall that the record breaking events characterized by $\sigma_k$'s are already 
correlated in the ``pure" random walk model, i.e., for $p=0$. How does a nonzero $p$ affect this correlation between the $\sigma_k$'s? Indeed we show 
a nonzero $p$ introduces additional {\it negative} correlations. More precisely, we first define the connected correlation function 
\begin{equation}\label{corr_funct1}
C_p(m_1 ,m_2)=
\langle\sigma_{m_1} \sigma_{m_2}\rangle -\langle\sigma_{m_1}\rangle\langle\sigma_{m_2}\rangle \;,
\end{equation}
where the subscript `$p$' indicates the nonzero staying probability. We show indeed that 
\bea \label{ineq}
C_p(m_1 ,m_2) < C_0(m_1 ,m_2) \quad, \quad {\rm for \; all}\; m_1, m_2 \quad {\rm and \; all}\quad 0< p \leq 1 \;.
\eea
Thus a nonzero $p$ in the jump distribution in Eq. (\ref{jump_distribution}) of the RW model induces additional anti-correlations
between the record-breaking events (see Fig. \ref{figure1}). As a consequence of these anti-correlations, the fluctuations of the record number up to step $N$, characterized
by its variance, get suppressed. The best way to visualise this effect is to study the 
Fano factor 
\bea \label{FNp}
F_N(p) = \frac{V_N(p)}{\langle R_N\rangle(p)}
\eea
as a function of increasing $p$, for fixed but large $N$. In the limit $p\to 0$, we have seen in Eq. (\ref{Fano_RW}) that $F_N(p=0) \equiv F_N \sim O(\sqrt{N})$. On the other hand 
we will see that as $p \to 1$, the Fano factor $F_N(p\to 1) \sim O(1)$. As $p$ increases, $F_N(p)$ decreases monotonically. In fact, the anti-correlations between the $\sigma_m$'s induced 
by a nonzero $p$ suppress both the mean $\langle R_N\rangle(p)$  as well as the variance $V_N(p)$ (see Figs. \ref{FigRp} and \ref{FigVp} respectively), but the variance gets suppressed more than the mean. Consequently, the Fano factor also decreases with increasing $p$ (see Fig. \ref{FigFp}).}

\end{itemize}

  At this point, it might be interesting to ask how these results are affected when also $f_0(\eta)$ ceases to be continuous. A natural choice would be to take for $f_0(\eta)$ a symmetric discontinuous distribution corresponding to a discrete space -- or `lattice' -- random walk (with $\eta\ne 0$). Results recently obtained in\ \cite{MMS2020} suggest that: (i) both Eq.~(\ref{ineq}) and a decreasing Fano factor with increasing $p$ are also obtained in the case of lattice random walks; but (ii) the results are no longer universal. Technically, this is due to the fact that results for lattice walks differ from the ones for continuous walks by the expression of $Z_p(s)$ only (see Eq.~(\ref{gSA:th2})), with a $f_0$-dependent $Z_p(s)$ in the lattice case (unlike the continuous case). Because of this lack of universality, calculations for lattice random walks can get much more cumbersome than their counterparts for continuous random walks, without yielding fundamentally different results. So, for simplicity, we will not consider lattice random walks further on in this paper.

The rest of the paper is organized as follows. In Section \ref{correlation}, we compute exactly, for the discrete-time model, the two-time correlation between the record events characterized by the binary variables $\sigma_m$'s and show that a nonzero staying probability $p$ reduces the connected correlation function as $p$ increases. In Section \ref{sec:exact}, we compute exactly, again for the discrete-time model, the statistics
of the number of records $R_N$ in $N$ steps: the mean (Section \ref{sec:exact} A), the variance (Section \ref{sec:exact} B) and the Fano factor (Section \ref{sec:exact} C), for arbitrary $N$ and arbitrary $0\leq p \leq 1$. We show, for any fixed $0\leq p \leq 1$, that these results are universal for any $N$, i.e. independent of the jump distribution $f_0(\eta)$. In Section \ref{sec:cont}, we consider the continuous-time scaling limit where $N \to \infty$, $p\to 1$ with the product $t = N\,(1-p)$ fixed.  In this limit, we compute the universal scaling functions associated with the mean, the variance and the Fano factor of the number of records. We also perform numerical simulations which show an excellent agreement with our analytical predictions. Finally, we conclude with a summary and some perspectives in Section \ref{sec:conclusion}. In Appendix \ref{app1}, we provide a physical interpretation of the formula for the survival probability $q_p(m)$ for a nonzero $p$.

%
%%%%%%%%%%%%%%%%%%%%
%
\section{Correlation between record events: exact universal expression}\label{correlation}

We start with the random walk sequence in (\ref{def_RW}), starting from $x_0=0$, where the jumps at each
step are independently drawn from the distribution $f(\eta)$ as in Eq. \eqref{jump_distribution}. This jump
distribution has two components: a delta function part at $\eta=0$ with weight $p$ and, with weight $1-p$, it
has a continuous and symmetric distribution $f_0(\eta)$. In this section, we compute the correlation
fonction between the record events characterized by the binary variables introduced before, namely,
\begin{equation}\label{sigma_funct}
\sigma_m=\left\lbrace
\begin{array}{ll}
1&{\rm if\ a\ record\ happens\ at\ step}\ m, \\
0&{\rm otherwise}\;,
\end{array}\right.
\end{equation}
for $m\ge 1$, and $\sigma_0=1$ (by convention, the initial position is counted as a record).
The total number of records up to step $N$ can then be expressed as a sum over $\sigma_m$'s 
\bea\label{eq:RN_sigma}
R_N = \sum_{m=0}^N \sigma_m \;.
\eea
Taking the average on both sides of (\ref{eq:RN_sigma}) we get
\bea\label{eq:RN_sigma_av}
\langle R_N\rangle(p)  = \sum_{m=0}^N \langle \sigma_m \rangle\;,
\eea
where $\langle \sigma_m \rangle$ is just the probability that a record happens at step $m$. Similarly, the second moment
of $R_N$ is given by
\bea\label{RNsq}
\langle R_N^2 \rangle(p) = \sum_{m_1=0}^N \sum_{m_2=0}^N \langle \sigma_{m_1} \sigma_{m_2}\rangle \;,
\eea
and the variance of $R_N$ can be expressed as
\bea \label{eq:varRN_sigma}
V_N(p) = \langle R_N^2 \rangle(p) - (\langle R_N \rangle(p))^2 = \sum_{m_1=1}^N  \sum_{m_2=1}^N C_p(m_1,m_2) \;,
\eea
where $C_p(m_1,m_2)$ is the connected correlation function of the $\sigma_m$'s defined in Eq. (\ref{corr_funct1}) and where we have used $C_p(0,m_2)=C_p(m_1,0)=0$ (as a consequence of $\sigma_0=1$). Thus to compute the mean and the variance
of $R_N$, we need to know the one-point and the two-point correlation functions of the $\sigma_m$'s.
It turns out to be convenient to separate the diagonal (i.e., $m_1=m_2$) and the off-diagonal (i.e., $m_1 \neq m_2$) parts. For the diagonal part, we use the identity $\sigma_m^2 = \sigma_m$ valid for any binary $(0,1)$ variable.  For the off-diagonal part, we can use the symmetry that the correlation function is invariant under the exchange $m_1 \leftrightarrow m_2$. Then it is straightforward to see that 
\bea \label{RNsq2}
\langle R_N^2 \rangle(p)= -\sum_{m=0}^N \langle \sigma_m \rangle + 2 \sum_{m_2=0}^N \sum_{m_1=0}^{m_2} \langle \sigma_{m_1} \sigma_{m_2}\rangle \;.
\eea

Consider first the one-point function $\langle \sigma_m \rangle$, denoting the probability that a record occurs at 
step $m$. This quantity is simply related to the survival probability of the walk defined as 
\bea \label{def_qm}
q_p(m)={\rm Prob}\, (x_1>0,\, x_2>0,\cdots ,\, x_m>0\vert x_0=0) \quad, \quad {\rm for} \quad m \geq 1 \;,
\eea
and $q_p(0)=1$. Indeed, 
\begin{eqnarray}\label{mean_sigma}
\langle\sigma_m\rangle &=&{\rm Prob}\, (``{\rm a\ record\ happens\ at\ step}\ m") = q_p(m) \;.
\end{eqnarray}
This relation can be understood as follows. In order that a record happens at step $m$,
we must have $x_m > \{x_0=0, x_1, \cdots, x_{m-1}\}$. Therefore, if we shift the origin of space
to the value $x_m$ and reverse the time, this event, using the symmetric nature of the walk,  
is precisely the survival probability $q_p(m)$ in Eq. (\ref{def_qm}). Consequently the mean number of records is
given by
\bea \label{av_qm}
\langle R_N \rangle(p) = \sum_{m=0}^N q_p(m) \;.
\eea

Similarly, the two-time correlation function can be expressed for $m_2 \geq m_1$ as  
\begin{eqnarray}\label{mean_sigma_sigma}
\langle\sigma_{m_1}\sigma_{m_2}\rangle
&=&{\rm Prob}\, (``{\rm records\ happen\ at\ steps}\ m_1\ {\rm and}\ m_2") = q_p(m_1)q_p(m_2-m_1) \;,
\end{eqnarray}
where we used the Markov property of the walk which makes the two intervals $[0,m_1]$ and $[m_1,m_2]$ statistically independent. 
Consequently the second moment of $R_N$ in Eq. (\ref{RNsq2}) can be expressed in terms of $q_p(m)$ 
\bea \label{RN_sq3}
\langle R_N^2 \rangle(p) = - \sum_{m=0}^N q_p(m) + 2 \sum_{m_2=0}^N \sum_{m_1=0}^{m_2} q_p(m_1) q_p(m_2-m_1)  \;.
\eea
Hence the variance of $R_N$ in Eq. (\ref{eq:varRN_sigma}) can also be expressed in terms of $q_p(m)$
\begin{equation} \label{RN_sq4}
V_N(p) = - \sum_{m=0}^N q_p(m) + 2 \sum_{m_2=0}^N \sum_{m_1=0}^{m_2} q_p(m_1) q_p(m_2-m_1)  - \left( \sum_{m=0}^N q_p(m)\right)^2\;.
\end{equation}

Finally, the connected correlation function can also be expressed in terms of only $q_p(m)$ 
\begin{equation}\label{corr_funct2}
C_p(m_1 ,m_2)= \langle \sigma_{m_1} \sigma_{m_2} \rangle - \langle \sigma_{m_1} \rangle \langle \sigma_{m_2} \rangle = 
q_p(m_1)\, \left\lbrack q_p(m_2-m_1)-q_p(m_2)\right\rbrack \;, \; m_2 \geq m_1 \;.
\end{equation}

Thus both the mean (\ref{av_qm}) and the variance (\ref{RN_sq4}) of $R_N$, as well as the connected correlation function $C_p(m_1 ,m_2)$ in (\ref{corr_funct2}) 
can all be expressed in terms of the single observable $q_p(m)$,
i.e., the survival probability for a random walk with a jump distribution $f(\eta) = p \delta(\eta) + (1-p) f_0(\eta)$
where $0\leq p \leq 1$ and $f_0(\eta)$ is symmetric and continuous. Fortunately, $q_p(m)$ can be computed using the generalised
Sparre Andersen theorem which states \cite{SA1954}
\bea \label{gSA:th}
Q_p(s) = \sum_{m\geq 0} q_p(m)\, s^m = \exp{\left[\sum_{n\geq 1} \frac{s^n}{n} \, {\rm Prob}(x_n<0) \right]} \;.
\eea
This result is highly nontrivial as it relates a history-dependent property (the survival probability) to an observable
which is local in time, namely the probability that the position of the walker at a given step $n$ is strictly negative. To compute the right
hand side of Eq. (\ref{gSA:th}) we use the symmetry property of the walk, i.e.  ${\rm Prob}(x_n<0)= {\rm Prob}(x_n>0)$. Furthermore, using 
the normalization at step $n$, we get 
\bea \label{norm0}
2 \, {\rm Prob}(x_n<0) + {\rm Prob}(x_n=0) = 1 \;,
\eea
giving 
\bea \label{norm}
{\rm Prob}(x_n<0) = \frac{1-{\rm Prob}(x_n=0)}{2} \;.
\eea 
Substituting this result on the right hand side of Eq. (\ref{gSA:th}) and using $\sum_{n \geq 1} s^n/n = - \ln(1-s)$, we get
\bea \label{gSA:th2}
Q_p(s) =\frac{1}{Z_p(s)\, \sqrt{1-s}} \quad, \quad {\rm where} \quad Z_p(s)=\exp\left\lbrack\frac{1}{2}\sum_{n=1}^{\infty}\frac{s^n}{n}\, {\rm Prob}\, (x_n=0) \right\rbrack \;.
\eea
We still need to compute the probability ${\rm Prob}\, (x_n=0)$. Given the jump distribution $f(\eta) = p \delta(\eta) + (1-p) f_0(\eta)$, we see that the walker arrives exactly at the origin 
after step $n$ if and only if it never quits the origin up to step $n$. Because if it does, then it will never come back {\it exactly} to the origin by continuous jumps drawn from $f_0(\eta)$. 
Since the probability to stay at the origin up step $n$ is simply $p^n$, we get 
\bea \label{Px0}
{\rm Prob}\, (x_n=0) = p^n \;.
\eea
Substituting this result in Eq. (\ref{gSA:th2}) we then have
\bea \label{gSA:th3}
Q_p(s) = \sum_{m\geq 0} q_p(m)\,s^m = \frac{\sqrt{1-s\,p}}{\sqrt{1-s}} \;.
\eea
Note that for $p=0$ this reduces to the well-known Sparre Andersen result $Q_0(s) = 1/\sqrt{1-s}$. Furthermore, even 
for a nonzero $p$, the result in Eq. (\ref{gSA:th3}) is completely universal, i.e., independent of $f_0(\eta)$. 

To extract explicitly $q_p(m)$ from Eq. (\ref{gSA:th3}), we use the power series expansions
$$
\sqrt{1-x}=\sum_{n\ge 0}(-1)^n\binom{1/2}{n}\, x^n,
$$
and
$$
\frac{1}{\sqrt{1-x}}=\sum_{n\geq 0}\binom{2n}{n} 2^{-2n}\, x^n = \sum_{n \geq 0} (-1)^n \binom{-1/2}{n}\, x^n,
$$
on the right-hand side of Eq.\ (\ref{gSA:th3}). Identifying the powers of $s$ on both sides, we get
\begin{eqnarray}\label{qgt0}
q_p(m)&=&(-1)^m\sum_{k=0}^{m}\binom{1/2}{k}\, \binom{-1/2}{m-k}\, p^k = (-1)^m\binom{-1/2}{m}\, {}_{2}F_{1}\left(-\frac{1}{2},-m\, ;\, \frac{1}{2}-m\, ;\, p\right) \nonumber \\
&=& \binom{2m}{m}2^{-2m}\,  {}_{2}F_{1}\left(-\frac{1}{2},-m\, ;\, \frac{1}{2}-m\, ;\, p\right)
\end{eqnarray}
where ${}_{2}F_{1}$ is the standard hypergeometric series \cite{AS}. Note that in the limit $p \to 0$, using ${}_{2}F_{1}(a,b; c;z=0)=1$, the result in Eq.~(\ref{qgt0}) reduces to the well known Sparre Andersen result $q_0(m) = \binom{2m}{m}2^{-2m}$. In Appendix\ \ref{app1} we give a physical interpretation of this formula for $q_p(m)$ in Eq. (\ref{qgt0}). 
Note that this expression for $q_p(m)$ is universal for all $m$ (and not just for large $m$), i.e. independent of $f_0(\eta)$. Interestingly, this formula (\ref{qgt0}) is very similar to the expression obtained
for the survival probability in a discrete-time persistent random walk model \cite{LM2020} [see Eq. (8) there], although the reason behind this similarity remains unclear. 

Substituting this formula for $q_p(m)$ in Eq. (\ref{corr_funct2}) gives an explicit formula for the correlation function $C_p(m_1,m_2)$, which thus is also universal. This formula involves complicated hypergeometric series so we do not display it explicitly. We have computed the difference 
\bea \label{diff}
\Delta_p(m_1, m_2) = C_p(m_1, m_2) -C_0(m_1, m_2)
\eea
and found, using Mathematica, that $\Delta_p(m_1,m_2)<0$, for all $0<p\leq 1$ and all $m_1, m_2 \geq 0$. We could verify analytically that  $\Delta_p(m_1,m_2)<0$ in the two limits $p\to 0$ and $p\to 1$. 
In addition, this can be verified analytically for all $p$ when $m_1$ and $m_2$ are both large (see below). Proving rigorously the general inequality $\Delta_p(m_1,m_2)<0$ for all $p>0$ and arbitrary $m_1, m_2 \geq 0$ seems challenging. However, one can provide a physical justification of this property (see the discussion at the end of this Section). This result thus shows that a nonzero $p$ introduces additional anti-correlations between the record breaking events $\sigma_m$'s. Later, we will see that these excess anti-correlations also suppress the variance of the record number $R_N$.

\vspace*{0.5cm}
\noindent{\it Asymptotic properties of $C_p(m_1, m_2)$.} While the expression of the correlation function $C_p(m_1,m_2)$, using Eqs. (\ref{corr_funct2}) and (\ref{qgt0}), 
 is explicit for finite $m_1$ and $m_2$, it is a bit cumbersome. Hence we now study the asymptotic behavior of $C_p(m_1, m_2)$ when both $m_1$ and $m_2$ are large. This expression simplifies considerably in this asymptotic limit.

To perform the asymptotic analysis of $C_p(m_1, m_2)$, we need to investigate the large $m$ behavior of the survival probability $q_p(m)$. This is most conveniently done using the explicit generating function $Q_p(s)$ in Eq. (\ref{gSA:th3}). To extract the large $m$ limit of $q_p(m)$ we need to analyse the $s \to 1$ limit of $Q_p(s)$. We see immediately that in this limit $Q_p(s) \simeq \sqrt{1-p}/\sqrt{1-s}$. Hence inverting the generating function gives
\begin{equation}\label{qgt0_largen}
q_p(m)\simeq\sqrt{\frac{1-p}{\pi m}}\ \ \ \ \ (m\to +\infty) \;.
\end{equation}
This asymptotic behavior can also be obtained from the representation in Eq. (\ref{qgt0}). Upon substituting this asymptotic behavior of $q_p(m)$ in Eq. (\ref{corr_funct2}) we get
\begin{equation}\label{corr_funct_asym}
C_p(m_1,m_2)\simeq
\frac{(1-p)}{\pi}\, \frac{1}{\sqrt{m_1}}\, \left(\frac{1}{\sqrt{m_2-m_1}}-\frac{1}{\sqrt{m_2}}\right) \;.
\end{equation}
Note that this result holds when both $m_1$ and $m_2$ are large and also their difference $m_2-m_1$ is large, while $p$ is kept fixed. Indeed, by computing the difference $\Delta_p(m_1,m_2)$ in Eq. (\ref{diff}) we see that 
\bea \label{diff_explicit}
\Delta_p(m_1, m_2) \simeq - \frac{p}{\pi} \left(\frac{1}{\sqrt{m_2-m_1}}-\frac{1}{\sqrt{m_2}}\right) \;.
\eea
Since $1/\sqrt{m_2-m_1}>1/\sqrt{m_2}$, this shows manifestly that $\Delta_p(m_1,m_2)<0$ for all $0<p\leq 1$.  

One can also investigate another scaling limit when $p \to 1$ while $m_1$ and $m_2$ are fixed. In this limit, $q_p(m)$, for fixed $m$, can be extracted again from the generating
function in Eq. (\ref{gSA:th3}). Setting $p = 1-\epsilon$ in Eq. (\ref{gSA:th3}) and expanding in powers of $\epsilon$ we get $q_p(0) = 1 $ and for $m\geq 1$
\begin{equation}\label{qgt0_pto1}
q_p(m)=\frac{(1-p)}{2}-\frac{(m-1)\, (1-p)^2}{8}+O\left((1-p)^3\right)\ \ \ \ \ (p\to 1) \;.
\end{equation}
Using this result in Eq.\ (\ref{corr_funct2}) one finds that in the $p\to 1$ limit with fixed $m_1\leq m_2$,
\begin{equation}\label{corr_funct_pto1}
C_p(m_1,m_2)\simeq
\frac{m_1\, (1-p)^3}{16}.
\end{equation}
It then follows again that $\Delta_p(m_1,m_2)<0$ for all $m_1, m_2 > 0$.

\begin{figure}
\centering
\includegraphics[width = 0.7\linewidth]{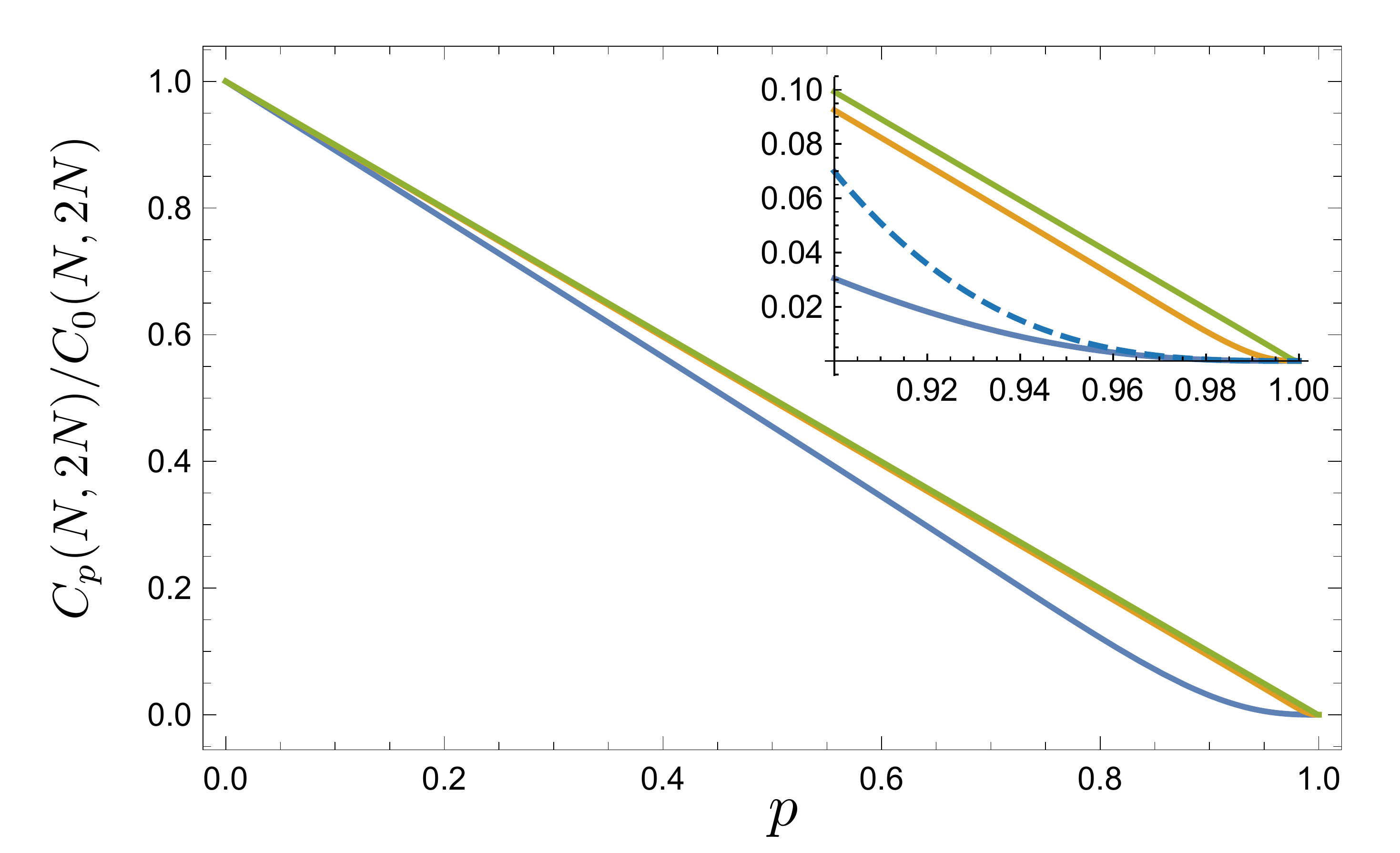}
\caption{Plots of $C_p(N,2N)/C_0(N,2N)$ as a function of $p$ for $N=10$ (blue), $N=100$ (orange), and $N=1000$ (green). Inset: enlargement of the same plots in the domain $0.9\le p\le 1$. The dashed line in the inset corresponds to the asymptotic expression in Eq.\ (\ref{corr_funct_pto1}) with $m_1=10$ (to be compared with the blue line).}\label{figure1}
\end{figure}

To illustrate the effect of a nonzero $p$ on the record correlation, we show in Fig. \ref{figure1} three plots of $C_p(N,2N)/C_0(N,2N)$ as a function of $p$ for $N=10$ (blue), $N=100$ (orange), and $N=1000$ (green). It can be seen that for $N=100$ and $N=1000$, the results are quasi indistinguishable from $1-p$, in agreement with the asymptotic behavior in Eq.\ (\ref{corr_funct_asym}). The inset is an enlargement of the same plots in the domain $0.9\le p\le 1$. Except for small values of $N=O(1)$, the asymptotic behavior in Eq.\ (\ref{corr_funct_pto1}) applies only for $p$ extremely close to $p=1$ (see the dashed line in Fig.\ \ref{figure1} for $N=10$).

Our results so far demonstrate that switching on the staying probability $p>0$ induces a reduction of the correlation function between record events. In order to better understand the physical reason behind this reduction it may be useful to rewrite $C_p(m_1,m_2)$ as
\begin{equation}\label{corr_funct3}
C_p(m_1,m_2)=q_p(m_1)q_p(m_2)\, g_p(m_1,m_2),
\end{equation}
where
\begin{eqnarray}\label{gcorr_funct}
g_p(m_1,m_2)
&=&\frac{{\rm Prob}\, ({\rm a\ record\ happens\ at\ step}\ m_2\ \vert\ {\rm a\ record\ happens\ at\ step}\ m_1)}
{{\rm Prob}\, ({\rm a\ record\ happens\ at\ step}\ m_2)} -1 \nonumber \\
&=&\frac{q_p(m_2-m_1)}{q_p(m_2)}-1.
\end{eqnarray}
The behavior of $C_p(m_1,m_2)$ is thus determined by the ones of $q_p(m_{1})$, $q_p(m_{2})$  and $g_p(m_1,m_2)$. For a nonzero staying probability $p>0$, stretches of walk where the walker stays in place get inserted between sections where she/he moves (and where records happen). As a result, the time between two given records is increased by the number of steps where the walker does not move in between, leading to a rarefaction of records (i.e. less records in a given time interval). This rarefaction of records translates into a reduction of $q_p(m)$, as can be seen in Eqs.\ (\ref{qgt0_largen}) and\ (\ref{qgt0_pto1}). As for the behavior of $g_p(m_1,m_2)$, it depends on the limit one considers. For fixed $p<1$ and  large $m_2$ and $m_2-m_1$, Eqs.\ (\ref{qgt0_largen}) and\ (\ref{gcorr_funct}) yield
\begin{equation}\label{gcorr_funct_asym}
g_p(m_1,m_2)\sim\sqrt{\frac{m_2}{m_2-m_1}} -1>0,
\end{equation}
which means that record events remain (positively) correlated in this limit, whatever the (fixed) value of $p$. In this case, the reduction of $C_p(m_1,m_2)$ observed in Eq.\ (\ref{corr_funct_asym}) for a non zero staying probability $p>0$ is due to the reduction of the factor $q_p(m_1)q_p(m_2)$ on the right-hand side of Eq.~(\ref{corr_funct3}), {\it not} to $g_p(m_1,m_2)$ which does not depend on $p$ for $m_{1}$ and $m_2$ large enough. In other words, for fixed $0<p<1$ and  large $m_2$ and $m_2-m_1$, the reduction of $C_p(m_1,m_2)$ must be attributed to the rarefaction of records which remain correlated, rather than to a loss of correlation between record events (which would correspond to a reduction of $g_p(m_1,m_2)$). The situation is different if one considers the limit $p\to 1$ at fixed $m_1$ and $m_2$. In this case, Eqs.\ (\ref{qgt0_pto1}) and\ (\ref{gcorr_funct}) give $q_p(m_1)q_p(m_2)\sim (1-p)^2/4$ and
\begin{equation}\label{gcorr_funct_pto1}
g_p(m_1,m_2)\sim\frac{m_1\, (1-p)}{4}\to 0 ,
\end{equation}
which means that record events tend to decorrelate as $p\to 1$. Thus, in this limit, the reduction of $C_p(m_1,m_2)$ observed in Eq.\ (\ref{corr_funct_pto1}) is due to both the rarefaction of records, i.e. the reduction of $q_p(m_1)q_p(m_2)$, by a factor $\sim (1-p)^2$, and to a loss of correlation between record events, i.e. a reduction of $g_p(m_1,m_2)$, by a factor $\sim (1-p)$.

\section{Exact statistics of records for arbitrary $\bm{0\leq p \leq 1}$}\label{sec:exact}

In this Section, we compute the mean, the variance and the Fano factor for the number of records $R_N$ explicitly for all $N$ and arbitrary $p$.

\subsection{Average number of records: exact universal expression}\label{average}
\begin{figure}[t]
\includegraphics[angle=90,width = 0.7\linewidth]{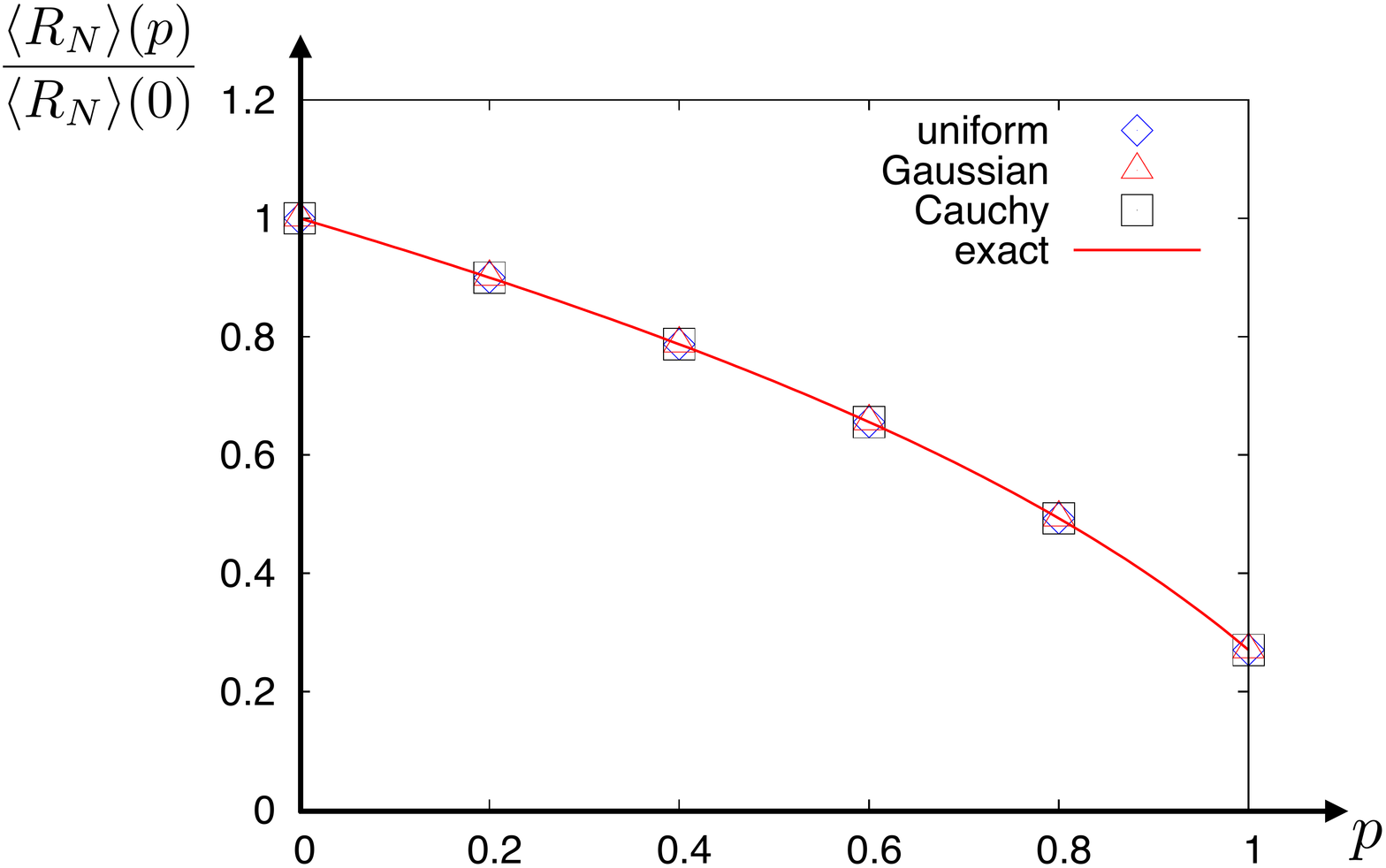}
\caption{Plot of the ratio $\frac{\langle R_N \rangle(p) }{\langle R_N \rangle(0) }$ vs $p$ for $N=10$. The symbols correspond to numerical simulations of random walks (\ref{def_RW}) with different jump distributions (uniform, Gaussian and Cauchy distributions) while the solid line corresponds to the exact result in Eq. (\ref{ratio_Rp}). The numerical results are clearly independent of the jump distributions and are in perfect agreement with our analytical prediction.}\label{FigRp}
\end{figure}

First, we determine the average number of records $\langle R_N \rangle$ by computing its generating function. Multiplying Eq. (\ref{av_qm}) on both sides by $s^N$ and summing over $N$ from $0$ to $\infty$ gives
\bea \label{gene_funct_ERn}
\sum_{N \geq 0} \langle R_N\rangle(p)\,s^N = \frac{1}{1-s} Q_p(s) =  \frac{\sqrt{1-sp}}{(1-s)^{3/2}} \;,
\eea
where $Q_p(s)$ is the generating function of $q_p(m)$ defined in Eq. (\ref{gSA:th3}). 
Using the power series expansions
\bea \label{dl1}
\sqrt{1-x}=\sum_{n\ge 0}(-1)^n\binom{1/2}{n}\, x^n,
\eea
and
\bea \label{dl2}
\frac{1}{(1-x)^{3/2}}=\sum_{n\ge 0}(-1)^n\binom{-3/2}{n}\, x^n,
\eea
on the right-hand side of Eq.\ (\ref{gene_funct_ERn}) and identifying the powers of $s$, one gets
\begin{eqnarray}\label{ERn}
\langle R_N \rangle(p) &=&(-1)^N\sum_{m=0}^{N}\binom{1/2}{m}\, \binom{-3/2}{N-m}\, p^m = (-1)^N\binom{-3/2}{N}\, {}_{2}F_{1}\left(-\frac{1}{2},-N\, ;\, -\frac{1}{2}-N\, ;\, p\right) \nonumber \\
&=& (2N+1) \binom{2N}{N}2^{-2N} \,{}_{2}F_{1}\left(-\frac{1}{2},-N\, ;\, -\frac{1}{2}-N\, ;\, p\right) \;.
\end{eqnarray}
The result in Eq.~(\ref{ERn}) is exact, valid for all $N$ and $p$. Moreover, for fixed $p$ and $N$, it is also completely universal, i.e., independent
of the continuous part of the jump distribution $f_0(\eta)$ in Eq. (\ref{jump_distribution}). Note that in the limit $p \to 0$, using ${}_{2}F_{1}(a,b; c;z=0)=1$, the result in Eq.~(\ref{ERn}) coincides with the previously known result given in Eq.~(\ref{av_RN_rw}). In the opposite limit $p=1$ one can check, using properties of the hypergeometric series, that 
\bea \label{Rnp1}
\langle R_N \rangle(p=1) = 1 \;.
\eea
This is expected since, for $p=1$, the walker does not move from the origin and hence the initial record at $N=0$ remains the only record. 
The ratio of the mean number of records at finite $p$ and at $p=0$ is given by
\bea \label{ratio_Rp}
\frac{\langle R_N \rangle(p) }{\langle R_N \rangle(0) } = {}_{2}F_{1}\left(-\frac{1}{2},-N\, ;\, -\frac{1}{2}-N\, ;\, p\right) \;.
\eea
As a function of $p$, for fixed $N$, this ratio decreases monotonically and is strictly less than $1$ for all $p>0$. A plot of this ratio as function of $p$ is shown in Fig. \ref{FigRp} for $N=10$ together with a comparison with numerical simulations (a qualitatively similar behaviour is observed for other values of $N$). Thus a nonzero staying probability $p$ suppresses the average number of records.

\vspace*{0.5cm}
\noindent{\it Asymptotic behavior of $\langle R_N \rangle(p)$.} By analysing the generating function in Eq. (\ref{gene_funct_ERn}) near $s=1$ we find, to leading order for large $N$ but fixed $p<1$, 
\begin{equation}\label{ERn_largen}
\langle R_N \rangle(p)= 2\, \sqrt{\frac{(1-p)\, N}{\pi}} + O(1/N^{1/2}) \, , \ \ \ \quad {\rm as} \quad N\to +\infty \;.
\end{equation}
In contrast, for fixed $N$ and $p \to 1$ limit, we get
\begin{eqnarray}\label{ERn_pto1}
\langle R_N \rangle(p)&=&1+\frac{N\, (1-p)}{2}-\frac{N\, (N-1)\, (1-p)^2}{16} \nonumber \\
&+&\frac{N\ (2-3N+N^2)\, (1-p)^3}{96}+O\left((1-p)^4\right) \;, \quad {\rm as} \quad \ \ \ \ p\to 1\;.
\end{eqnarray}
%
%
%%%%%
%
\subsection{Variance of the number of records}\label{variance}

\begin{figure}[t]
\includegraphics[width = 0.7 \linewidth]{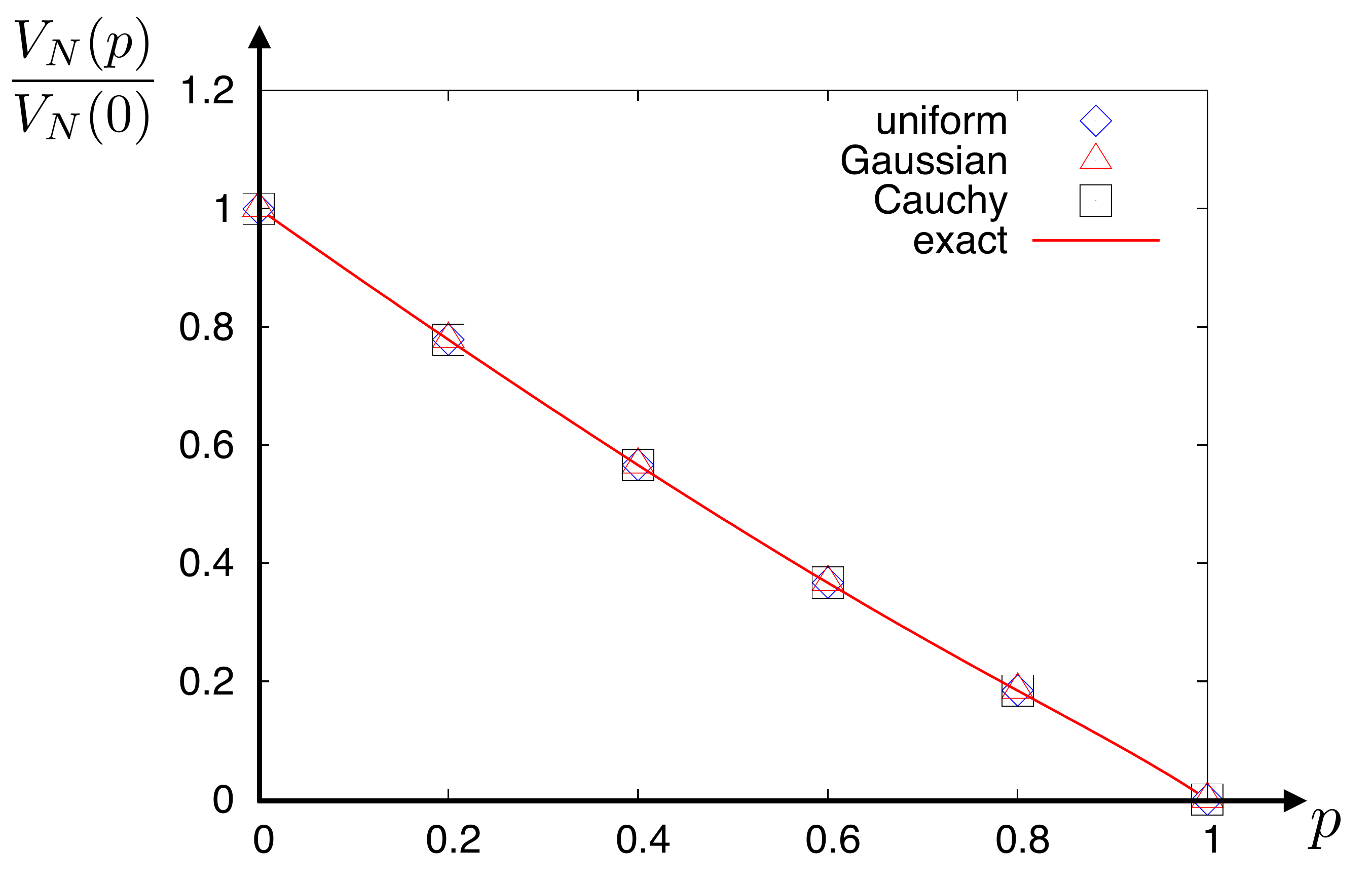}
\caption{Plot of the ratio $\frac{V_N(p)}{V_N(0)}$ vs $p$ for $N=10$. The symbols correspond to numerical simulations of random walks (\ref{def_RW}) with different jump distributions (uniform, Gaussian and Cauchy distributions) while the solid line corresponds to the exact result in Eq. (\ref{VarRn}). The numerical results are clearly independent of the jump distributions and are in perfect agreement with our analytical prediction.}\label{FigVp}
\end{figure}

Our starting point is the expression of the second moment of $R_N$ in Eq. (\ref{RN_sq3}). We multiply by $s^N$ on both sides of Eq. (\ref{RN_sq3}) and sum over $N$ from $0$ to $\infty$.
This gives, using the convolution structure of the double sum
\bea \label{RNsq_GF}
\sum_{N \geq 0} \langle R_N^2 \rangle(p) s^N = \frac{1}{1-s}\left(2\,Q_p^2(s) - Q_p(s) \right) \;,
\eea
where $Q_p(s)$ is given in Eq. (\ref{gSA:th3}). Using the explicit expression of $Q_p(s)$ in  Eq. (\ref{gSA:th3}), we get
\begin{eqnarray}\label{gene_funct_ERn2}
\sum_{N\ge 0} \langle R_N^2\rangle(p)\, s^N &=&\frac{2\, (1-sp)}{(1-s)^2}-\frac{\sqrt{1-sp}}{(1-s)^{3/2}} \;.
\end{eqnarray}  
We use the representation
\bea \label{rep1}
\frac{2\, (1-sp)}{(1-s)^2}=2\sum_{n\ge 0}\lbrack (1-p)\, n+1\rbrack\, s^n \;,
\eea
and then identify the powers of $s$ on both sides of Eq.\ (\ref{gene_funct_ERn2}). Using further Eq. (\ref{gene_funct_ERn})
finally gives a very simple formula
\begin{equation}\label{ERn2}
\langle R_N^2 \rangle(p) =2\lbrack (1-p)\, N+1\rbrack - \langle R_N\rangle(p) \;.
\end{equation}
It follows immediately that
\begin{equation}\label{VarRn}
V_N(p)=2\lbrack (1-p)\, N+1\rbrack -\langle R_N\rangle(p) \lbrack \langle R_N \rangle(p)+1\rbrack \;.
\end{equation}
The result in Eq.~(\ref{VarRn}) is exact, valid for all $N$ and $p$. Moreover, for fixed $p$ and $N$, since the mean $\langle R_N\rangle(p)$ is universal for all
$N$, the variance $V_N(p)$ in Eq. (\ref{VarRn}) is also completely universal, i.e., independent
of the continuous part of the jump distribution $f_0(\eta)$ in Eq. (\ref{jump_distribution}). In the limit $p \to 0$, it reduces to the known formula $V_N(0) = 2(N+1)-\langle R_N \rangle(\langle R_N \rangle + 1)$ where $\langle R_N \rangle \equiv \langle R_N \rangle(p=0)$ is given in Eq. (\ref{av_RN_rw}). In the opposite limit $p = 1$, Eq. (\ref{VarRn}) gives $V_N(1) = 0$ which is expected since $R_N = 1$ with probability one in this case. As a function of $p$, for fixed $N$, the ratio of the variance at finite $p$ and the one at $p=0$ is strictly less than $1$ for all $p>0$ and, in addition, decreases monotonically with increasing $p$. A plot of this ratio as a function of $p$ is shown in Fig. \ref{FigVp} for a representative value $N=10$, together with a comparison with numerical simulations. Thus a nonzero staying probability $p$ suppresses also the variance of the record number.

\vspace*{0.5cm}
\noindent{\it Asymptotic behavior of $V_N(p)$.} We first consider the case for fixed $p<1$ and large $N$. 
In this case, substituting the large $N$ behavior of $\langle R_N \rangle(p)$ from Eq. (\ref{ERn_largen}) in Eq. (\ref{VarRn}), we get
\bea\label{Var_largeN}
V_N(p) = 2(1-p)\left(1 - \frac{2}{\pi} \right)\, N + O(\sqrt{N}) \; \quad {\rm as} \quad N \to \infty \;.
\eea
In contrast, for fixed $N$ and in the limit $p \to 1$, one obtains using Eqs. (\ref{ERn_pto1}) and (\ref{VarRn})
\bea \label{Var_p1}
V_N(p) &=& \frac{N}{2}(1-p) - \frac{1}{16}N(N+3)(1-p)^2 \nonumber \\
&+& \frac{1}{32}N(N^2+N-2)(1-p)^3 + O((1-p)^4) \; \quad {\rm as} \quad p \to 1 \;.
\eea 

\subsection{Fano factor}\label{sec:Fano}

\begin{figure}[t]
\includegraphics[width = 0.7 \linewidth]{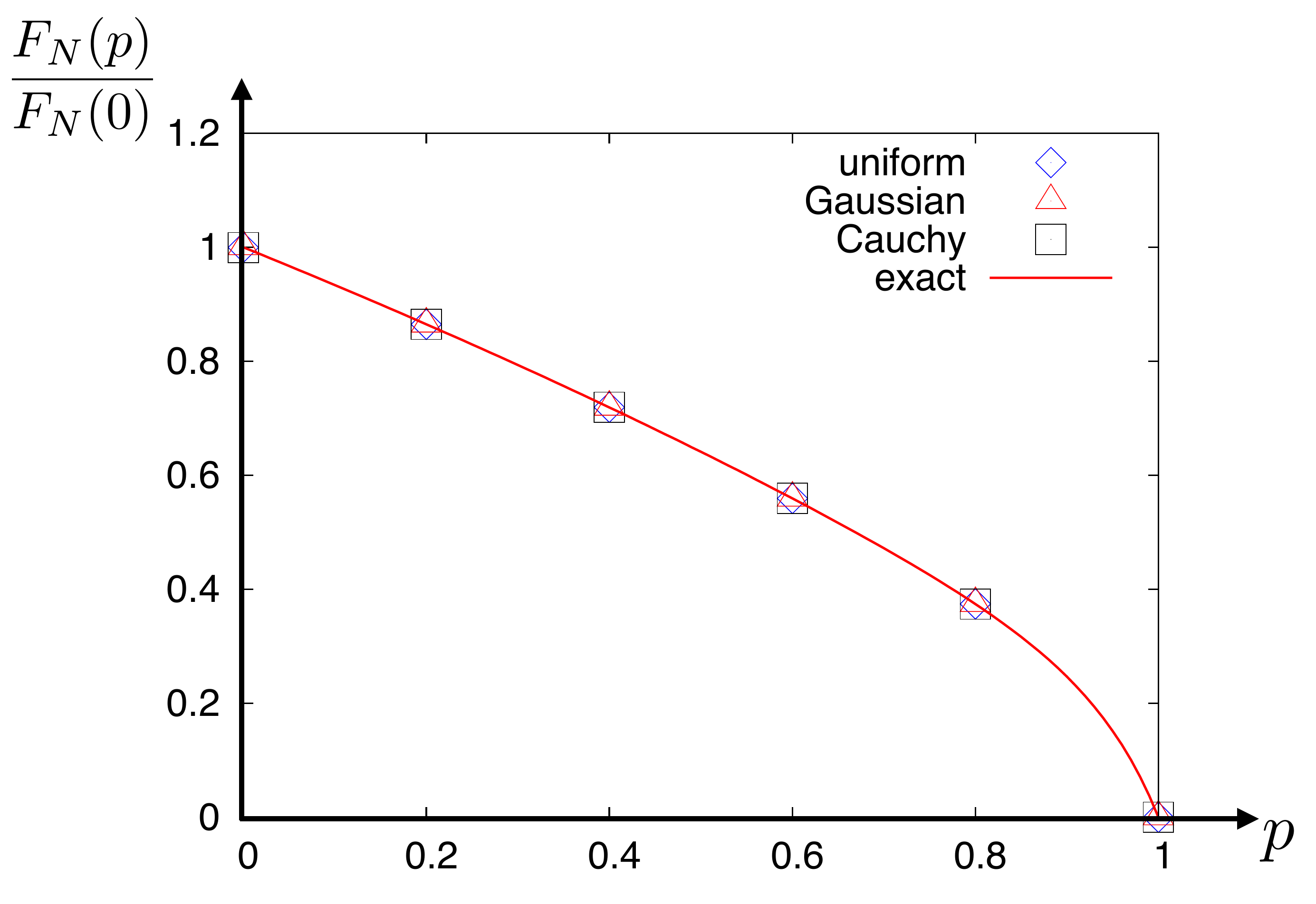}
\caption{Plot of the ratio $\frac{F_N(p)}{F_N(0)}$ vs $p$ for $N=10$. The symbols correspond to numerical simulations of random walks (\ref{def_RW}) with different jump distributions (uniform, Gaussian and Cauchy distributions) while the solid line corresponds to the exact result in Eq. (\ref{Fano_exp}). The numerical results are clearly independent of the jump distributions and are in perfect agreement with our analytical prediction.}\label{FigFp}
\end{figure}

So far, we have seen that the effect of a nonzero staying probability $p$ is to suppress both the mean as well as the variance of the number of records $R_N$ up to step
$N$. It is then interesting to know the relative suppression, which is measured by the Fano factor defined in Eq. (\ref{FNp}). Dividing Eq. (\ref{VarRn}) by $\langle R_N\rangle(p)$
we obtain
\bea \label{Fano_exp}
F_N(p) = \frac{2\left[ N\,(1-p)+1\right]}{\langle R_N\rangle(p)} - \langle R_N\rangle(p) - 1 \;.
\eea
Note that $F_N(p)$ is also universal for all $N$ and $p$ since both the mean and the variance are universal. We find from Eq. (\ref{Fano_exp}) that
$F_N(p)$, for fixed $N$, decreases monotonically with increasing $p$. Thus, a nonzero staying probability suppresses the variance more than the mean. 
As in the cases of the mean and the variance, we plot the ratio $F_N(p)/F_N(0)$ as a function of $p$ in Fig. \ref{FigFp} for a representative value of $N=10$, together with a comparison with numerical simulations. 
Clearly, this ratio is strictly less than $1$ for all $p>0$ and also it decreases with increasing $p$.  

\vspace*{0.5cm}
\noindent{\it Asymptotic behavior of $F_N(p)$.} For fixed $p<1$, and large $N$, we find from Eq. (\ref{Fano_exp}), using~(\ref{ERn_largen}),  
\bea \label{Fano_largen}
F_N(p) = \left(1 - \frac{2}{\pi} \right) \sqrt{(1-p)\, \pi\,N} - 1 + O(1/\sqrt{N})  \; \quad {\rm as} \quad N \to \infty \;.
\eea
In the limit where $N$ is fixed but $p \to 1$, we get from Eq. (\ref{Fano_exp}), using~(\ref{ERn_pto1})
\bea\label{Fanop1}
F_N(p)&=& \frac{N}{2}(1-p) - \frac{N(5N+3)}{16}(1-p)^2 \nonumber \\
&+& \frac{N}{32}(7N^2 + 3N-2)(1-p)^3  + O((1-p)^4) \; \quad {\rm as} \quad p \to 1 \;.
\eea

\section{Correlation and Fano factor in the continuous time limit}\label{sec:cont}

In Sections\ \ref{correlation} and\ \ref{sec:exact} we have studied the two limits $N\to +\infty$ at fixed $p$ and $p\to 1$ at fixed $N$. The corresponding results clearly show that these two limits do not commute, which suggests the existence of a scaling regime describing the crossover between the leading asymptotic results for large $N$ at fixed $p$ and small $(1-p)$ at fixed $N$. As we will see below, this scaling regime is defined by the limits $N\to +\infty$ and $p\to 1$ keeping $(1-p)\, N=t$ fixed. In fact, as discussed in the introduction, in this scaling limit, the model reduces to the continuous time random walk (CTRW) model with exponential waiting-time distribution. This is also the IR model mentioned before \cite{MDMS2020b}. In fact, the record statistics in the CTRW model with arbitrary waiting time and jump distribution was studied in Ref. \cite{Sanjib2011}. In the IR model, when the waiting time is purely exponential, the mean number of records in a fixed time interval $[0,t]$ was computed explicitly and was found to be universal at all times $t$, i.e. independent of the jump distribution as long as it is symmetric and continuous \cite{MDMS2020b}. 

In this Section, by taking the scaling limit ($p \to 1$, $N \to \infty$ with $t = N\,(1-p)$ fixed) of our exact discrete-time results valid for all $p$ and $N$, we show that we do recover the known result for the mean of the IR model. In addition, we also compute the variance and the Fano factor in this scaling limit. Most importantly, we show that the anti-correlations between the record events persist even in this scaling limit. As in the case for fixed $p$ and $N$, in the scaling limit, the anti-correlations also effectively reduce the mean as well as the variance of the number of records.

%%%%%
%
\subsection{Correlation between record events}\label{correlation_scaling}

To compute the correlations between record events in the scaling limit, we start from the expression of
the connected correlation function $C_p(m_1, m_2)$ in Eq. (\ref{corr_funct2}), valid for arbitrary $m_2 \geq m_1$.
Therefore, we need to compute the survival probability $q_p(m)$ in the scaling limit $p \to 1$, $m \to \infty$, with
$t = (1-p)\,m$ fixed. 

\begin{figure}[t]
\includegraphics[width = 0.6\linewidth]{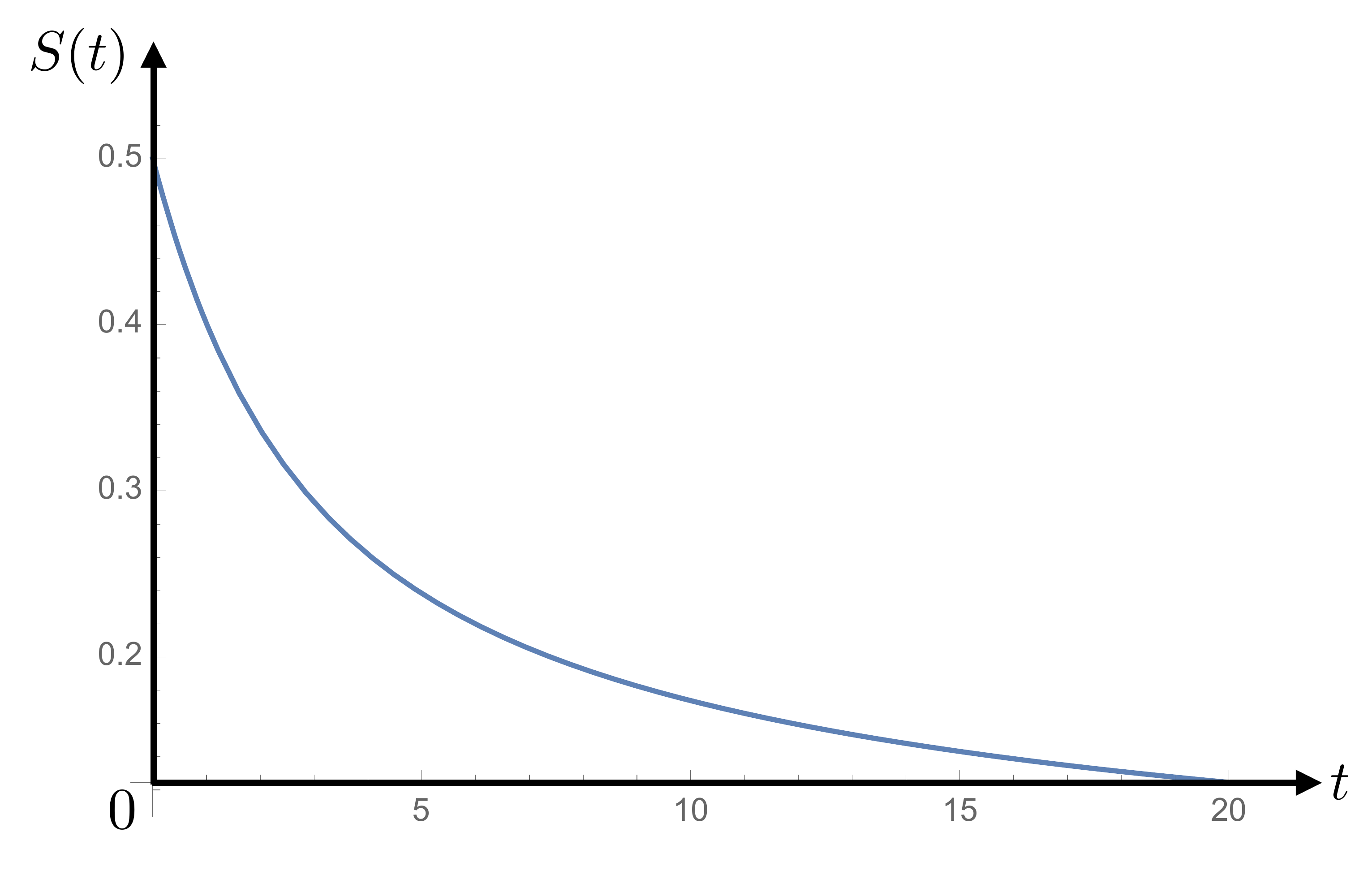}
\caption{Plot of $S(t)$ vs $t$ as given in Eq. (\ref{scaling_functqgt0}).}\label{Fig_Soft}
\end{figure}

Inverting the generating function in Eq. (\ref{gSA:th3}) using Cauchy's theorem \cite{Hen}, one gets
\begin{equation}\label{qgt0_integral}
q_p(m)=\frac{1}{2i\pi}\oint\frac{\sqrt{1-sp}}{s^{m+1}\sqrt{1-s}}\, ds \;,
\end{equation}
where the integral is along a contour encircling the origin in the complex $s$-plane. To take the scaling limit, we first
make a change of variable $s=\exp(-\lambda/m)$ in Eq.\ (\ref{qgt0_integral}). In the $m\to +\infty$ limit, the dominant contribution to the integral
comes from the vicinity of $s=1$. Hence, to leading order, we can write
\begin{equation}\label{qgt0_integral_asym}
q_p(m)\simeq\frac{1}{2i\pi\,m}
\int_{\mathcal{L}}\frac{\sqrt{\lambda +(1-p)\, m}}{\sqrt{\lambda}}\, {\rm e}^{\lambda}d\lambda
\ \ \ \ \ (m \to +\infty ,\ p\to 1) \;,
\end{equation}
where $\mathcal{L}$ is a Bromwich contour which runs along the imaginary axis in the complex $\lambda$-plane. Performing then the integral on the right-hand side of Eq.\ (\ref{qgt0_integral_asym}), one gets the universal scaling form
\begin{equation}\label{qgt0_scaling}
q_p(m)\simeq (1-p)\, S\left\lbrack (1-p)\, m\right\rbrack ,
\end{equation}
valid for $m\to +\infty$, $p\to 1$, and fixed $(1-p)\, m=t$, with the scaling function
\begin{equation}\label{scaling_functqgt0}
S(t)=\frac{1}{2}
\, \left\lbrack I_0\left(\frac{t}{2}\right)+I_1\left(\frac{t}{2}\right)\right\rbrack\, {\rm e}^{-t/2},
\end{equation}
where $I_\nu(t)$ is the modified Bessel function of order $\nu$. A plot of this function $S(t)$ is given in Fig. \ref{Fig_Soft}. From the large and small argument behaviors of $I_\nu(t)$, one has
\begin{equation}\label{scaling_functqgt0_asym}
S(t)\simeq\left\lbrace
\begin{array}{ll}
\dfrac{1}{2}-\dfrac{t}{8}\quad, \quad &{\rm as} \quad t\to 0 \\
\quad & \quad \\
\dfrac{1}{\sqrt{\pi t}}\quad, \quad &{\rm as} \quad t\to  +\infty\;. \\
\end{array}
\right.
\end{equation}

It can be checked that Eq.\ (\ref{qgt0_scaling}), with the asymptotic behaviors of $S(t)$ in Eq.\ (\ref{scaling_functqgt0_asym}), coincides with Eq.\ (\ref{qgt0_largen}) for $(1-p)\, m\gg 1$ and with Eq.\ (\ref{qgt0_pto1}) for $(1-p)\, m\ll 1$, as it should be.

Injecting Eq.\ (\ref{qgt0_scaling}) onto the right-hand side of Eq.\ (\ref{corr_funct2}), one gets the universal scaling form for the correlation function,
\begin{equation}\label{corr_funct_scaling}
C_p(m_1,m_2)\simeq (1-p)^2 \mathcal{C}\left\lbrack (1-p)\, m_1,(1-p)\, m_2\right\rbrack ,
\end{equation}
valid for $m_1,\, m_2\to +\infty$, $p\to 1$, and fixed $(1-p)\, m_1=t_1$ and $(1-p)\, m_2=t_2$, with the scaling function for $t_2\geq t_1$
\begin{equation}\label{scaling_funct_corr}
\mathcal{C}(t_1,t_2)=
S(t_1)\, \left\lbrack S(t_2-t_1)-S(t_2)\right\rbrack \;,
\end{equation}
where $S(t)$ is given in Eq. (\ref{scaling_functqgt0}). 

%%%%%
%
\subsection{Average number of records, variance and the Fano factor}\label{Fano_scaling}

\begin{figure}[t]
\includegraphics[width = 0.5\linewidth]{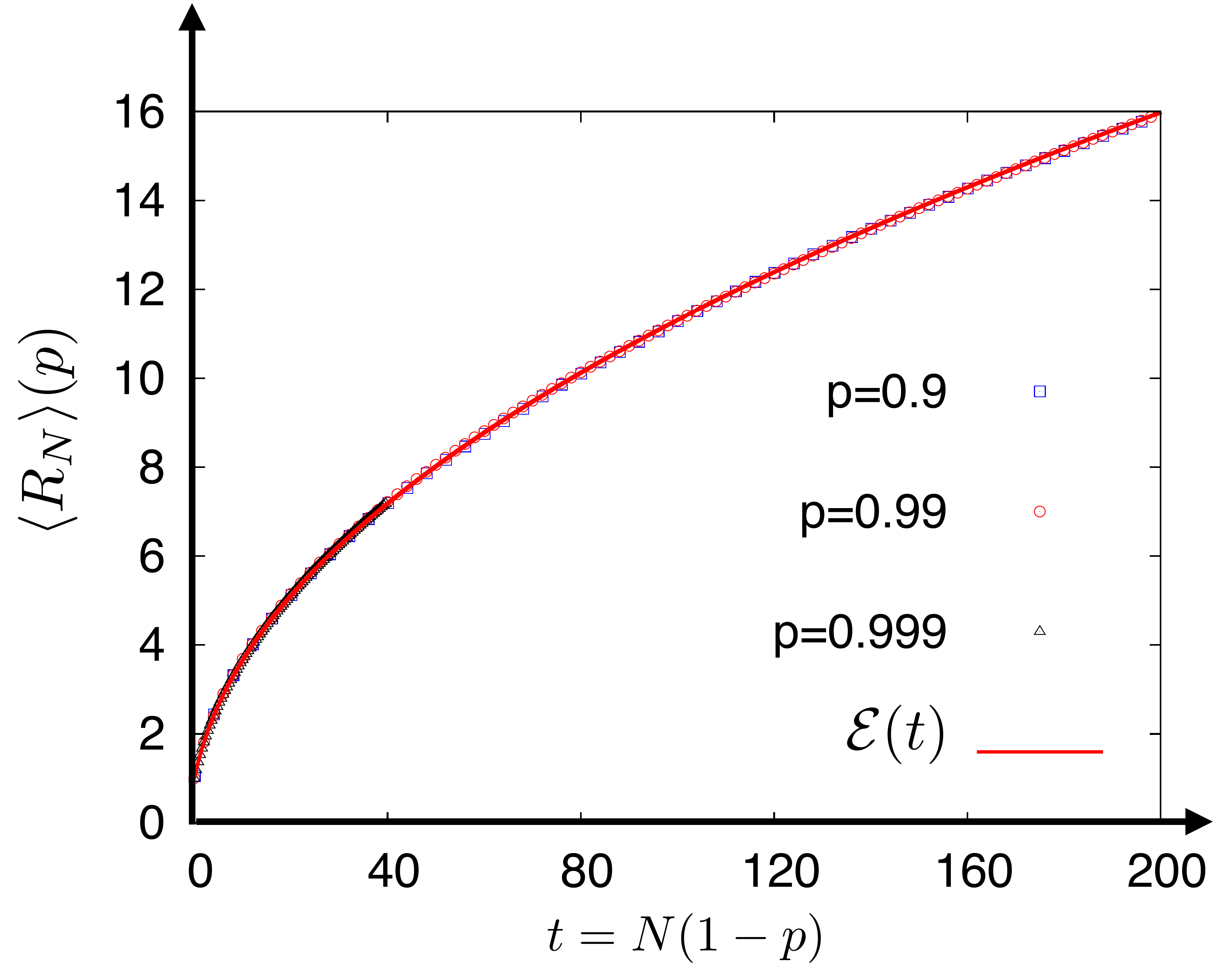}
\caption{Plot of $\langle R_N\rangle(p)$ vs the scaling variable $t = N(1-p)$ for three representative values of $p$ close to $1$ and $N=40000$. The symbols correspond to the simulations while the solid red line is the theoretical scaling function ${\cal E}(t)$ given in Eq. (\ref{scaling_functE}). The agreement is excellent as evident from the fact the symbols are almost indistinguishable from the solid line.}\label{Fig:av_scaling}
\end{figure}

\vspace*{0.5cm}
\noindent{\it Average number of records}. By inverting the generating function in Eq.\ (\ref{gene_funct_ERn}) we get
\begin{equation}\label{ERn_integral}
\langle R_N \rangle(p)=\frac{1}{2i\pi}\oint\frac{\sqrt{1-sp}}{s^{N+1}(1-s)^{3/2}}\, ds \;.
\end{equation}
As before, in the scaling limit, setting $s=\exp(-\lambda/N)$ and noting that the integral is dominated by the vicinity of $s=1$, one can convert this contour integral into a Browmich integral in the complex $\lambda$-plane
\begin{equation}\label{ERn_integral_asym}
\langle R_N \rangle(p)\simeq\frac{1}{2i\pi}
\int_{\mathcal{L}}\frac{\sqrt{\lambda +(1-p)\, N}}{\lambda^{3/2}}\, {\rm e}^{\lambda}d\lambda
\ \ \ \ \ (N\to +\infty ,\ p\to 1) \;.
\end{equation}
Performing the Bromwich integral on the right-hand side of Eq.\ (\ref{ERn_integral_asym}), one gets the universal scaling form
\begin{equation}\label{ERn_scaling}
\langle R_N \rangle(p)\simeq\mathcal{E}\left\lbrack (1-p)\, N\right\rbrack ,
\end{equation}
valid for $N\to +\infty$, $p\to 1$, and fixed $(1-p)\, N=t$, with the scaling function
\begin{equation}\label{scaling_functE}
\mathcal{E}(t)=\left\lbrack (1+t)\, I_0\left(\frac{t}{2}\right)+t\, I_1\left(\frac{t}{2}\right)\right\rbrack
\, {\rm e}^{-t/2}.
\end{equation}
From the large and small argument behaviors of $I_\nu(t)$, we get
\begin{equation}\label{scaling_functE_asym}
\mathcal{E}(t)\simeq\left\lbrace
\begin{array}{ll}
1+\dfrac{t}{2} \quad, \quad &{\rm as} \quad t\to 0 \\
\quad & \quad \\
2\sqrt{\dfrac{t}{\pi}} \quad, \quad &{\rm as} \quad t\to \infty \;.
\end{array}
\right.
\end{equation}
One can check that the asymptotic behaviour as $t\to 0$ is consistent with Eq. (\ref{ERn_pto1}) in the limit $p \to 1$. 
On the other hand, the behavior as $t \to \infty$ is consistent with the large $N$ behavior given in Eq. (\ref{ERn_largen}). We note that the scaling function $\mathcal{E}(t)$ coincides with the average number of records in the continuous time IR model obtained, by a rather different method, in Ref. \cite{MDMS2020b}. This is expected since, as we argued before, our discrete-time model reduces to the continuous-time IR model in the scaling limit $N \to \infty$, $p \to 1$ with $t = N(1-p)$ fixed. By inspecting Eq. 
(\ref{scaling_functE}) and Eq.\ (\ref{scaling_functqgt0}), one finds an exact relation 
\begin{equation}\label{scaling_functs_relation}
\mathcal{E}(t)=1+\int_0^t S(\tau)\, d\tau \;.
\end{equation}
This relation is not surprising as it follows by substituting the scaling form for $q_p(m)$ in Eq.~(\ref{qgt0_scaling}) into the exact relation $\langle R_N\rangle(p)-1=\sum_{m=1}^n q_p(m)$ 
in Eq. (\ref{av_qm}).

\begin{figure}[t]
\includegraphics[width = 0.5\linewidth]{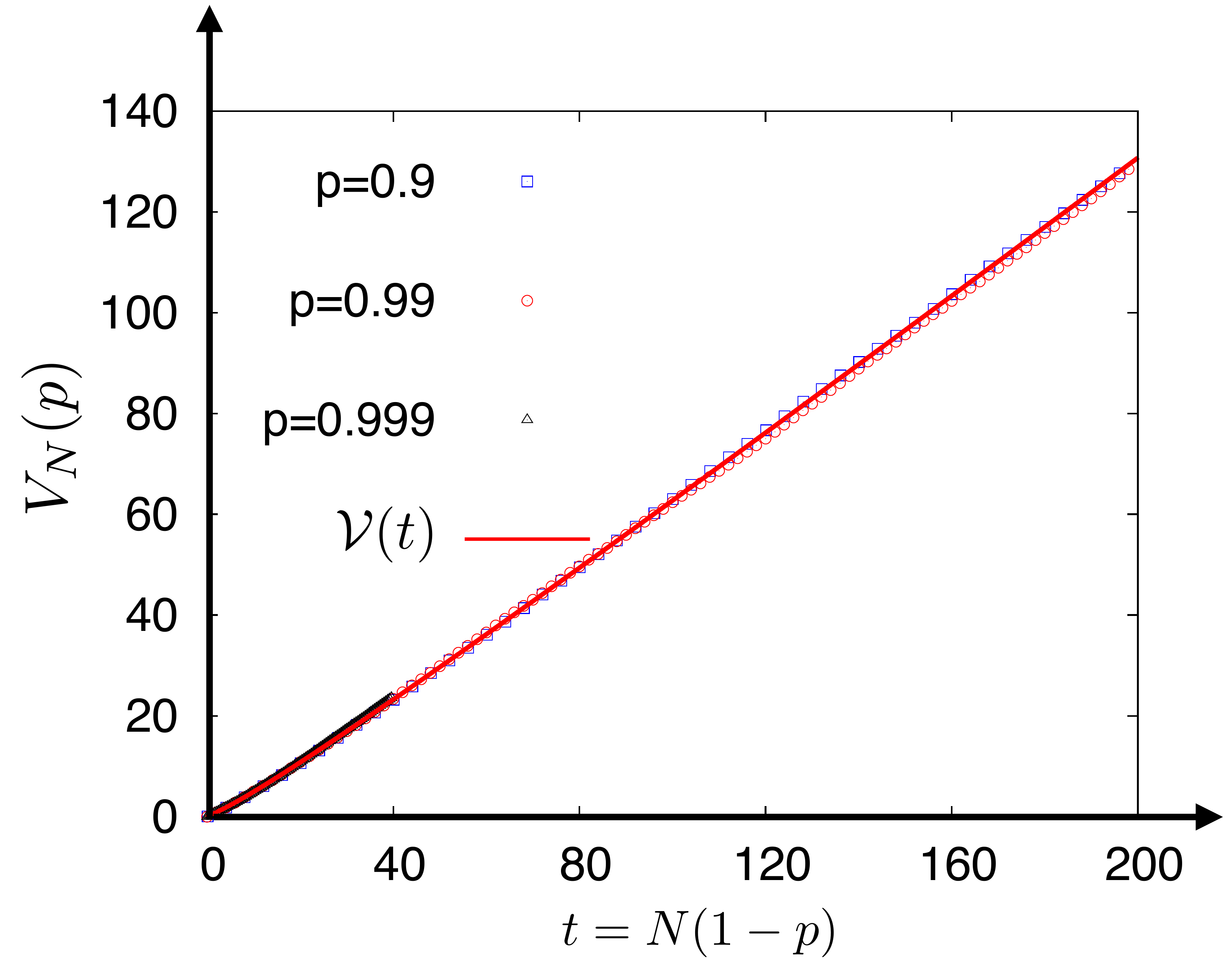}
\caption{Plot of $V_N(p)$ vs the scaling variable $t = N(1-p)$ for three representative values of $p$ close to $1$ and $N=40000$. The symbols correspond to the simulations while the solid red line is the theoretical scaling function ${\cal V}(t)$ given in Eq. (\ref{Voft}). The agreement is again excellent.}\label{Fig:var_scaling}
\end{figure}
\vspace*{0.5cm}
\noindent{\it Variance and the Fano factor}. We start with the exact result for $V_N(p)$ in Eq. (\ref{VarRn}) and inject the scaling form of $\langle R_N \rangle(p)$ in Eq. (\ref{ERn_scaling}). This gives the variance in the scaling limit as
\bea \label{VN_scaling}
V_N(p) \simeq {\cal V}(N\,(1-p)) \;,
\eea
where the scaling function ${\cal V}(t)$ is given by
\bea \label{Voft}
{\cal V}(t) = 2(t+1) - {\cal E}(t)\left({\cal E}(t)+1\right) \;,
\eea
with ${\cal E}(t)$ given in Eq. (\ref{scaling_functE}). The asymptotic behaviors of ${\cal V}(t)$ are given by
\begin{equation}\label{scaling_functV_asym}
\mathcal{V}(t)\simeq\left\lbrace
\begin{array}{ll}
\dfrac{t}{2} \quad, \quad &{\rm as} \quad t\to 0 \\
\quad & \quad \\
2\left(1-\dfrac{2}{\pi} \right)t \quad, \quad &{\rm as} \quad t\to \infty \;.
\end{array}
\right.
\end{equation}
These asymptotic behaviours are consistent with the two limiting behaviours given in Eqs.~(\ref{Var_p1}) and (\ref{Var_largeN}) respectively.

\begin{figure}[t]
\includegraphics[width = 0.5\linewidth]{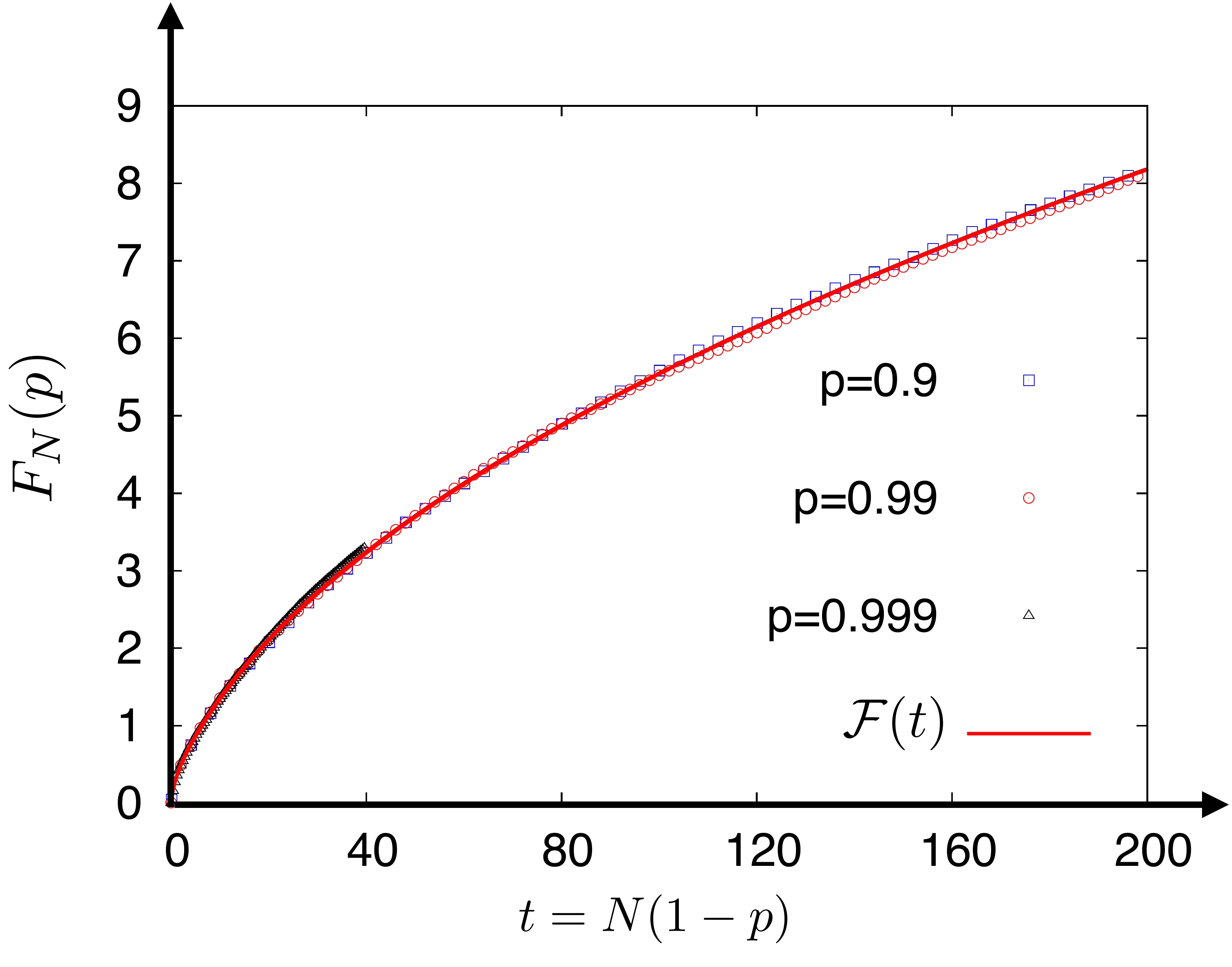}
\caption{Plot of $F_N(p)$ vs the scaling variable $t = N(1-p)$ for three representative values of $p$ close to $1$ and $N=40000$. The symbols correspond to the simulations while the solid red line is the theoretical scaling function ${\cal F}(t)$ given in Eq. (\ref{calFt}). The agreement between the simulations and the theoretical scaling function is very good.}\label{Fig:fano_scaling}
\end{figure}
Consequently, the Fano factor also has a scaling form 
\bea \label{scaling_F}
F_N(p)\simeq {\cal F}(N\,(1-p)) \;,
\eea
with the scaling function 
\bea \label{calFt}
{\cal F}(t) = \frac{2(t+1)}{{\cal E}(t)} - {\cal E}(t) - 1 \;,
\eea
with ${\cal E}(t)$ given in Eq. (\ref{scaling_functE}). Its asymptotic behaviors are given by 
\begin{equation}\label{scaling_functF_asym}
\mathcal{F}(t)\simeq\left\lbrace
\begin{array}{ll}
\dfrac{t}{2} \quad, \quad &{\rm as} \quad t\to 0 \\
\quad & \quad \\
\left(1 - \dfrac{2}{\pi} \right)\sqrt{\pi\, t} \quad, \quad &{\rm as} \quad t\to \infty \;.
\end{array}
\right.
\end{equation}
These asymptotic behaviours are consistent with the two limiting behaviours given in Eqs.~(\ref{Fanop1}) and (\ref{Fano_largen}) respectively. 

Finally, we have also performed numerical simulations in the scaling limit $p\to 1$, $N \to \infty$ with the product $t = N\, (1-p)$ fixed. Our numerical results are completely  
consistent with the scaling behaviors of the mean, the variance and the Fano factor given respectively in Eqs. (\ref{ERn_scaling}), (\ref{VN_scaling}) and (\ref{scaling_F}). In Figs. \ref{Fig:av_scaling},  \ref{Fig:var_scaling} and  \ref{Fig:fano_scaling} we compare our analytical
predictions for the scaling functions to numerical simulations, finding excellent agreements. Note that the observed slight bendings of the numerical curves (compared to the exact predictions) are
due to the fact that the parameters $p$ and $N$ are at the ``border'' of the scaling regime, which corresponds to $p \to 1$ and $N \to \infty$.

\section{Summary and conclusion}\label{sec:conclusion}

In this paper, we have studied the record statistics in a discrete-time random walk model on a line where the walker stays at a given position with a nonzero
probability $0\leq p \leq 1$, while with the complementary probability $1-p$, it jumps to a new position with a jump length drawn from a continuous and symmetric
distribution $f_0(\eta)$. We have shown that, for arbitrary $p$, the statistics of records up to step $N$ is completely universal, i.e., independent of $f_0(\eta)$ for
any $N$ (and not just for large $N$). In the limit $p \to 0$, this corresponds to the standard random walk model with continuous and symmetric jump
distribution \cite{MZ2008}. In the opposite limit, $p \to 1$, our model reduces to a continuous time random walk (CTRW) model with an exponential waiting-time distribution. The record statistics of 
the latter model was studied recently in the context of run and tumble processes in $d$-dimensions \cite{MDMS2020b}. In both limits, the record statistics was known to be universal and our
model, interpolating between these two limits, demonstrates that the universality with respect to $f_0(\eta)$ holds for all $0\leq p \leq 1$. 

One of the main messages of our paper is to elucidate the role of anti-correlations between the record-breaking events that are induced by a nonzero staying 
probability $p$. The role of such anti-correlations on record statistics was recently studied in the context of a rainfall precipitation time series with uncorrelated entries \cite{MBK2019}.
Our study is a generalization of this model to a time-series whose entries correspond to the positions of a random walk and hence are strongly correlated. In our model,
we have computed exactly the connected correlation function $C_p(m_1, m_2)$ of the record-breaking events at two times $m_1$ and $m_2$. One of our main results
is to show that the increment in the correlation function due to a nonzero $p$, $\Delta_p(m_1, m_2) = C_p(m_1, m_2)-C_0(m_1, m_2)$ is {\it negative} for all $p$, quantifying
the anti-correlations. We have shown that these anti-correlations reduce both the mean and the variance of the number of records as $p$ increases. However, it has a
more pronounced effect on the variance compared to the mean. As a result, the Fano factor (the ratio of the variance and the mean) also decreases with increasing $p$. In particular,
while in the $p \to 0$ limit the Fano factor scales with the number of steps $N$ as $O(\sqrt{N})$ for large $N$, it becomes of order $O(1)$ as $p \to 1$, signalling a drastic
reduction of the fluctuations of the record number with increasing $p$.

As mentioned above, our model is a discrete-time version of the continuous time ``Instantaneous Run'' model recently studied in the context of a run-and-tumble
process of active particles. In the context of active particles, another interesting continuous-time process is the so-called ``Instantaneous Tumble'' (IT) model where
a particle runs during an exponentially distributed random time, followed by an instantaneous change of direction known as ``tumbling''. Recently, a discrete-time
version of this model was studied in Ref. \cite{LM2020} for which the record statistics was also computed exactly and shown to be universal, i.e., independent of the jump
distribution as in our model here. However, the correlations between the record breaking events have not been studied in this IT model, and it would be interesting
to study the role of these correlations. 

In this paper, we have focused only on the mean, the variance and the two-time correlation function between the record-breaking events and shown them to be
universal for all $N$ and $p$, i.e., independent of the jump distribution $f_0(\eta)$. In fact, it is straightforward to carry out our analysis to higher moments of $R_N$ and 
higher order correlation functions between the record-breaking events. It is clear that these higher order observables are also going to be universal and it would be
interesting to compute them explicitly.   

Finally, there are other observables going beyond the statistics of the number of records. For instance, it would be interesting to study the statistics of the ages
of the records in this model -- an age of a record is the number of steps it remains a record before being broken by the next record \cite{review_records}. In  the limit $p \to 0$,
the age statistics has been studied extensively in this random walk model \cite{MZ2008,GMS2014}. It will be interesting to extend these studies to an nonzero $p$. In particular one may ask: 
how do the anti-correlations between the record-breaking events for a nonzero $p$ affect the age statistics of records?

\acknowledgments 
We thank Francesco Mori for useful discussions.

%\newpage

\appendix

\section{Interpretation of the formula giving $\bm{q_p(n)}$}\label{app1}

In the record statistics of the random walk model studied here, the basic building block is the survival property $q_p(m)$ defined in Eq. (\ref{def_qm}) and computed explicitly in Eq.\ (\ref{qgt0}). All other observables associated to the number of records, such as its mean and variance, can be expressed in terms of $q_p(m)$. In this Appendix, we show that the expression for $q_p(m)$ in Eq.\ (\ref{qgt0}) has an alternative representation which provides a nice physical interpretation. In a similar spirit to the one described for the IICD model in Ref. \cite{MBK2019}.

Using the relation {15.3.5} in\ \cite{AS} on the right-hand side of Eq.\ (\ref{qgt0}) and expanding the hypergeometric function in power series of its last argument, one gets
\begin{eqnarray}\label{qgt0bis}
q_p(m)&=&(-1)^m\binom{-1/2}{m}\, {}_{2}F_{1}\left(-\frac{1}{2},-m\, ;\, \frac{1}{2}-m\, ;\, p\right) \nonumber \\
&=&(p-1)^m\, \binom{-1/2}{m}\, 
{}_{2}F_{1}\left(-m,1-m\, ;\, \frac{1}{2}-m\, ;\, \frac{p}{p-1}\right) \nonumber \\
&=&(p-1)^m\sum_{k=0}^{m-1}\binom{-1/2}{m-k}\binom{m-1}{k}\, \left(\frac{p}{p-1}\right)^k \nonumber \\
&=&(1-p)\sum_{k=0}^{m-1}q_{p=0}(m-k)\, P_p(k),
\end{eqnarray}
with
\begin{equation}\label{conditional_qgt0}
q_{p=0}(m-k)=(-1)^{m-k}\, \binom{-1/2}{m-k},
\end{equation}
and
\begin{equation}\label{p_of_m}
P_p(k)=\binom{m-1}{k}p^k (1-p)^{m-k-1}.
\end{equation}
Equation\ (\ref{qgt0bis}) provides a simple interpretation of the formula giving $q_p(m)$ in Eq.\ (\ref{qgt0}). The reasoning goes as follows: by switching on the staying probability $p>0$, one allows stretches of walk where the walker does not move to get inserted between sections where she/he moves. Let $k$ be the total duration of such stretches, i.e., $k$ is the total number of steps where the walker stays in place. For a given $k$, it is clear that the survival probability reduces to $q_{p=0}(m-k)$ in Eq.~(\ref{conditional_qgt0}), i.e. the survival probability for a random walk with $p=0$ and $m-k$ steps (the $m-k$ remaining steps where the walker moves). Now, to get $q_p(m)$ it remains (i) to multiply by $(1-p)$, the probability that the walker moves at the first step, which is a necessary condition for a realization to contribute to $q_p(m)$ in Eq. (\ref{def_qm}) where the position of the walker needs to be strictly positive; (ii) to multiply by $P_p(k)$ in Eq.\ (\ref{p_of_m}), the probability that the walker stays in place $k$ times among the $m-1$ steps after the first one; and (iii) to sum over $k$. By doing so, one obtains the last equality on the right-hand side of Eq.\ (\ref{qgt0bis}), hence the formula\ (\ref{qgt0}).
%
%%%%%%%%%%
%
%
%%%%%%%%%%%%%%%%%%%%
%

%
\end{document}